\documentclass[10pt,journal,a4paper,oneside,twocolumn]{IEEEtran}

\usepackage{amsfonts}
\usepackage{mathrsfs}
\usepackage{graphicx}
\usepackage{cite}
\usepackage{citesort}
\usepackage{color}
\usepackage{psfrag}
\usepackage{subfigure}
\usepackage{amssymb}
\usepackage{epsfig}
\usepackage{pifont}
\usepackage{amsmath}
\usepackage[amsmath,thmmarks]{ntheorem}
\usepackage{array}
\usepackage{multicol}
\usepackage{algorithm}
\usepackage{algorithmic}
\allowdisplaybreaks[1]

\renewcommand{\baselinestretch}{0.932}

\newtheorem{theorem}{Theorem}

\newtheorem{lemma}{Lemma}
\makeatletter
\def\ifundefined{\@ifundefined}
\makeatother 

\begin{document}

\title{Fast Adaptive S-ALOHA Scheme for Event-driven M2M Communications}


\author{\IEEEauthorblockN{Huasen Wu\IEEEauthorrefmark{1}\IEEEauthorrefmark{2},
Chenxi Zhu\IEEEauthorrefmark{3}, Richard J. La\IEEEauthorrefmark{4},
Xin Liu\IEEEauthorrefmark{2}, and Youguang
Zhang\IEEEauthorrefmark{1}\\}
\IEEEauthorblockA{\IEEEauthorrefmark{1}School of Electronic and
Information Engineering, Beihang University, Beijing\\}
\IEEEauthorblockA{\IEEEauthorrefmark{2}Department of Computer Science, University of California, Davis\\}
\IEEEauthorblockA{\IEEEauthorrefmark{3}Mallard Creek Networks, 11452
Mallard Creek Trail, Fairfax, VA\\}
\IEEEauthorblockA{\IEEEauthorrefmark{4}Department of Electrical and
Computer Engineering, University of Maryland, College Park, MD}}


\maketitle 
%
%
%
%
%
%
%
%
%

\begin{abstract}
Supporting massive device transmission is challenging in Machine-to-Machine (M2M) communications. Particularly, in event-driven M2M communications, a large number of devices activate within a short period of time, which in turn causes high radio congestions and severe access delay. To address this issue, we propose a Fast Adaptive S-ALOHA (FASA) scheme for random access control of M2M communication systems with bursty traffic. Instead of the observation in a single slot, the statistics of consecutive idle and collision slots are used in FASA to accelerate the tracking process of network status which is critical for optimizing S-ALOHA systems. Using drift analysis, we design the FASA scheme such that the estimate of the backlogged devices converges fast to the true value. Furthermore, by examining the $T$-slot drifts, we
prove that the proposed FASA scheme is stable as long as the average arrival rate is smaller than $e^{-1}$, in the sense that the Markov chain derived from the scheme is geometrically ergodic. Simulation results demonstrate that the proposed FASA scheme outperforms traditional additive schemes such as PB-ALOHA and achieves near-optimal performance in reducing access delay. Moreover, compared to multiplicative schemes, FASA shows its robustness under heavy traffic load in addition to better delay performance.
\end{abstract}


\section{Introduction}\label{sec:introd}
\IEEEPARstart{M}{achine}-to-Machine (M2M) communication or
Machine-Type Communication
(MTC) is expected to be one of the major drivers of cellular networks \cite{Cisco2011TR} and has become one of the focuses in 3GPP \cite{3GPP2011TS22368,3GPP2010R2_100204,3GPP2010R2_104662}. Behind the proliferation of M2M communication, the congestion problems in M2M communication become a big concern. The reason is that the device density of M2M communication is much higher than that in traditional Human-to-Human (H2H) communication \cite{3GPP2011TS22368,3GPP2010R2_100204}. For example, it is expected in \cite{3GPP2010R2_100204} that 1,000
devices/km$^2$ are deployed for environment monitoring and control.
What is worse, in event-driven M2M applications, many devices may be
triggered almost simultaneously and attempt to access the base
station (BS) through the Random Access Channel (RACH)
\cite{Bertrand2009RA}. Such high burstiness can result in congestion
and increase access delay, which motivates our research.

Several efforts have been made in 3GPP to alleviate the radio congestion on RACH. Back to 3GPP LTE (Long Term Evolution), since the amount of exchanged data grows rapidly, congestion control has been addressed and access class barring (ACB) scheme has been proposed for overload protection \cite{3GPP2005TR23898}. In ACB scheme, devices are divided into several access classes. Before establishing a connection, the device is required to perform the ACB check and randomly transmit request packets with a probability broadcasted by the BS. Therefore, the traffic load can be reduced by choosing a small transmission probability. It is noticed that ACB scheme is a good candidate for congestion control in M2M applications, though modifications in accordance with the specific features of M2M applications is necessary \cite{3GPP2010R2_100182}. Therefore, in \cite{3GPP2010GP101378}, a two stage access control scheme, which consists of a primary level and a secondary level of access control barring, is introduced to provide prioritized M2M services based on their service attributes. Moreover, the authors in \cite{Lien&Chen2012ITWC} propose a cooperative ACB scheme for balancing traffic load among BSs in a heterogeneous multi-tier cellular network. With cooperations among BSs, the congestion level can be reduced and the access delay can be significantly improved. However, a key problem arising in implementing these schemes is how to estimate the number of active devices and optimize the transmission probability. This problem becomes even worse in event-driven M2M applications which is characterized with highly bursty traffic.

Essentially, the ACB scheme belongs to slotted-ALOHA (S-ALOHA) type schemes, which are widely applied for random access control. To address the instability issue of S-ALOHA \cite{Rosenkrantz1983InstabilityALOHA}, plenty of work has been done for deciding the protocol parameters and stabilizing the S-ALOHA system, which is briefly summarized in Section \ref{sec:related_work}. In these schemes, historical outcomes are applied to estimate the network status and adjust protocol parameters. Drift analysis is then used to design the schemes and prove their stability \cite{Kelly1985StochModels,Hajek1982DriftAnal,Tsitsiklis1987StabAnal}. However, these schemes usually rely on the assumption that the traffic can be modeled as a Poisson process and only apply the observation in the previous slot for estimating the number of backlogged devices. Due to the burstiness, this assumption cannot be justified in the context of M2M applications and the observation in a single slot is not satisfactory for adjusting the protocol parameters in time. Thus, we try to make full use of the information provided by the historical outcomes for improving the performance of access control under bursty traffic.

In this paper, we study adaptive S-ALOHA scheme for event-driven M2M communication and provide rigorous analysis about its stability. As our main contribution, we propose a Fast Adaptive S-ALOHA (FASA) scheme for the random access control of M2M devices. A key characteristic of FASA is that the access results in the past slots, in particular, consecutive idles or collisions, are collected and applied to estimate the number of backlogged devices. This enables the fast update of transmission probability under highly bursty traffic and thus reduce the access delay. Furthermore, we prove the stability of FASA under bursty traffic. This is accomplished by examining the $T$-slot drifts of the network status, which captures the memory property of FASA. Under interrupted Poisson
traffic model \cite{Kuczura1973InterruptedPoisson}, we show that the
$T$-slot drifts of FASA have the required properties for stabilizing
the scheme and the system is stable when the arrival rate is less
than $e^{-1}$. Numerical results demonstrate that using FASA scheme, the transmission probability of S-ALOHA can be adjusted in time and the access delay can be reduced to be very close to
the theoretical lower bound under highly bursty traffic.

The remainder of the paper is organized as follows. Section~\ref{sec:related_work} summarizes the related work. In
Section~\ref{sec:sys_model}, we present the system model, including
the bursty traffic model for the event-driven M2M communications. In
Section~\ref{sec:algorithm}, after analyzing the limitations of
traditional fixed step size policies, we propose the FASA scheme and design the parameters in the scheme based on drift analysis. Then in
Section~\ref{sec:stability}, we study the $T$-slot drifts of FASA and prove its
stability. In Section~\ref{sec:sim_res},
simulation results are presented to evaluate the performance of the
proposed scheme, compared with the theoretical optimal scheme in ideal case and two traditional adaptive schemes. Finally, we conclude our paper in
Section~\ref{sec:conclusions}.

\section{Related Work}\label{sec:related_work}

\subsubsection*{Radio Congestion Control in M2M communications}
Due to the high equipment density, the radio congestion of M2M applications is still an open problem. Besides the efforts in 3GPP \cite{3GPP2010R2_100182,3GPP2010GP101390,Lien&Chen2012ITWC}, there are a few publications addressing this issue. Since group-based feature appears in many M2M applications, some researchers propose hierarchical architectures, such as grouping scheme \cite{Jung2010Grouping} and relay schemes \cite{Kim2011CCIS,Andreev2011GlobeCom}, for alleviating the radio congestion on the RACH. In these architectures, group heads are selected for collecting the messages from the group members. However, efficient schemes for communications between the group head and members is required to reduce the access delay. At the same time, Adaptive Traffic Load Slotted Multiple Access Collision Avoidance (ATL S-MACA) mechanism proposed in \cite{Wang2010ICWITS} uses packet sensing and adaptive method to improve the access performance under high traffic load. However, the scheme is designed for Poisson traffic and is not suitable for event-driven M2M applications.

\subsubsection*{Adaptive S-ALOHA Schemes}
S-ALOHA is a fundamental scheme for ransom access control and the combinations with other techniques such as CDMA make it even more useful. However, the instability issue of S-ALOHA should be dealt with when being implemented in real networks.  Two typical classes of schemes, additive and multiplicative adaptive schemes, have been proposed for stabilizing S-ALOHA systems \cite{Cunningham1990ITC}. In these schemes, the estimate of the network status is updated in a additive or multiplicative manner, respectively, and the transmission probability is adjusted accordingly. However, as discussed in more detail later, traditional additive schemes such as Pseudo Bayesian ALOHA (PB-ALOHA) \cite{Rivest1987PBALOHA} estimate the number of backlogged devices based on the access result in the previous slot but cannot adjust the transmission probability in time under highly busty traffic, resulting in large access delay. On the other hand, because of the exponential increment in consecutive collision slots or decrement in consecutive idle slots, multiplicative schemes \cite{Hajek1982ITAC}, e.g., Q-Algorithm in
\cite{ISO18000_6} and its enhanced version Q$^+$-Algorithm in \cite{Lee2007QPlus}, can track the network status in a short period. However, the throughput suffers in these schemes due to the fluctuations in the estimation \cite{Hajek1982ITAC}. Therefore, we aim to design adaptive schemes that could adjust the protocol parameters fast under bursty traffic while retaining the same stable throughput as additive schemes. In our previous work \cite{Wu2011FASA}, we propose a preliminary version of FASA based on some intuitive approximations and show its desirable properties through numerical simulations. However, no rigorous analysis about the stability of FASA is presented in \cite{Wu2011FASA}.


\subsubsection*{Drift Analysis for stabilization of S-ALOHA}
Drift analysis is a theory for deducing the properties such as ergodicity of a sequence from its drift and is found useful in the design and analysis of adaptive S-ALOHA schemes \cite{Carleial1975BistableALOHA,Hajek1982DriftAnal,Gurcan2001ITVT}. The network status, which is represented by the number of backlogged devices and its estimate, could be viewed as a stochastic sequence. It is shown in \cite{Hajek1982DriftAnal} that when the drifts of the network status satisfies some criterions, the system is stable in the sense that the returning time can be bounded with high probability. Using the conclusion in \cite{Hajek1982DriftAnal}, the most related work \cite{Tsitsiklis1987StabAnal} studies the stability of PB-ALOHA scheme by defining a Lyapunov function to represent the network status and examining its drift. In all the work mentioned above, the schemes update the parameters based on the observation in the previous slot. Thus, the 1-slot drifts, i.e., drifts between two adjacent slots, are sufficient for studying the stability of the systems. When involving access results in multiple slots, however, the $T$-slot drifts are required to deal with the memory property of our scheme. Unlike 1-slot drifts, calculating $T$-slot drifts is non-trivial and we have to resort to approximations for obtaining their properties.


\section{System Model}\label{sec:sys_model}

In this section, we describe random access control procedure for M2M
communication as well as the traffic model, which will be used for
studying the stability and evaluating the performance of the
proposed scheme.

\subsection{S-ALOHA Based Random Access}
\begin{figure}[htbp]
\begin{center}
\includegraphics[angle = 0,width = 0.8\linewidth]{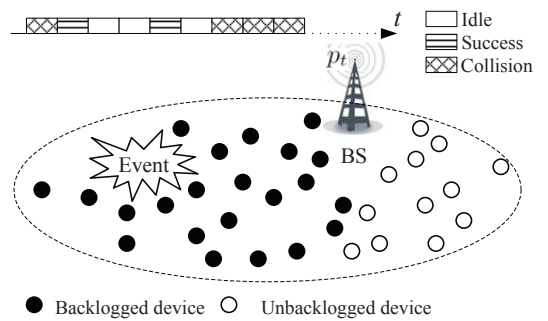}
\caption{S-ALOHA based access control of M2M communications}
\label{fig:network_struct}
\end{center}
\vspace{-0.5cm}
\end{figure}

We consider a cellular network based M2M communication system for
event detection. As shown in Fig.~\ref{fig:network_struct}, the system consists of a BS and a large number of
M2M devices. When an event is detected, certain number of devices
are triggered and attempt to access the BS by sending request
packets through a single RACH based on S-ALOHA scheme.

The time is divided into time slots, each of which is long enough to
transmit a request packet. Deferred first transmission (DFT) mode
\cite{Rivest1987PBALOHA} is assumed, in which a device with a new
request packet immediately goes to backlogged state. In the $t$th
slot, where $t\in \mathbb{Z}_+ := \{0,1,2, \ldots\}$, all backlogged
devices transmit packets with probability $p_t$, which is
broadcasted by the BS at the beginning of the slot. For the sake of
simplicity, we assume that the request packets generated by M2M
devices will eventually be transmitted successfully and a backlogged
device will not generate any new requests since the new coming data
can be transmitted as long as the device accesses the BS successfully.
Moreover, an ideal collision channel is assumed, where the
transmitted packet will be successfully received by the BS when no
other packets are being transmitted in the same slot.

The transmission probability $p_t$ is adjusted based on access
results in the past. Let $Z_t$ denote the access result in slot $t$,
and $Z_t = 0$, 1, or $c$ depending on whether zero, one, or more
than one request packets are transmitted on the RACH. At the end of
slot $t$, the BS decides the transmission probability for next slot
based on the sequence $\{Z_0,Z_1, \ldots, Z_t\}$, i.e.,
\begin{equation*} \label{eq:decide_tran_prob}
p_{t+1} = \Gamma_t(Z_0, Z_1, \ldots, Z_t).
\end{equation*}
The objective of the BS is to maximize the throughput and minimize
the access delay. It well known that, when $N_t \geq 1$, where $N_t$ is the number of
backlogged devices in slot $t$, using a
transmission probability $p_{t} = 1/N_t$ in slot $t$ maximizes the
throughput of the S-ALOHA system. However, the BS does not know $N_t$
and has to obtain its estimate $\hat{N}_t$ based on the access
results in the past.

\subsection{Traffic Model}
In order to capture the burstiness of event-driven M2M traffic,
instead of traditional Poisson process, the arrival process is
modeled as an {\it interrupted} Poisson process, which was suggested
by Hayward of Bell Laboratories for simulating overflow traffic
\cite{Kuczura1973InterruptedPoisson}.

Interrupted Poisson process can be viewed as a Poisson process
modulated by a random switch and will be discretized according to
the slotted structure of the scheme. Let $Y_t$ and $A_t$ respectively denote the number of events happening and the number of devices triggered in slot $t$. Assume that events happen
independently and identically in each slot and at most one event
happens in one slot, with $\theta$ being the happening probability. Hence, in each slot $t$, ${\rm Pr}(Y_t = 1) = \theta$ and ${\rm Pr}(Y_t = 0) = 1- \theta$. In addition, assume that the number of triggered devices follows Poisson distribution with mean $\lambda$ when an event happens, and no devices become active otherwise, i.e., $A_t \sim \mathcal
{P} (\lambda)$ when $Y_t = 1$ (ON-state), and $A_t = 0$ when $Y_t = 0$ (OFF-state). Therefore,
random variables $\{A_t: t\in \mathbb{Z}_+\}$ are independently and
identically distributed (i.i.d.) and the long term arrival rate can be
calculated as
\begin{equation} \label{eq:}
\bar{\lambda} =   {\rm Pr}(Y_t = 0)\cdot0 +  {\rm Pr}(Y_t = 1)\cdot
\lambda = \theta \lambda.
\end{equation}
In addition, the burstiness of the traffic is reflected by the
variance of $A_t$, which is given by
\begin{equation} \label{eq:}
\sigma_A^2 = \theta (\lambda +  \lambda^2) - (\theta \lambda)^2 =
\bar{\lambda}(1+\lambda-\bar{\lambda}).
\end{equation}

We note that the classic Poisson process is included as a special
case of this model when $\theta = 1$. We will design and analyze
adaptive S-ALOHA scheme based on this traffic model. Indeed, as will
be discussed later, the scheme proposed in this paper could be
stable under some other more general traffic models.

\section{Fast Adaptive S-ALOHA}\label{sec:algorithm}

The estimation of the number of backlogged devices plays an
important part in stabilizing and optimizing the S-ALOHA system. In
this section, using drift analysis, we first examine the limit of traditional
fixed step size estimation schemes.
Then, we propose and analyze a fast adaptive scheme, referred to as
Fast Adaptive S-ALOHA.

\subsection{Drift Analysis of Fixed Step size Estimation Schemes}
Many schemes with fixed step size have been proposed in the
literature to estimate the number of backlogged devices
\cite{Cunningham1990ITC}. A unified framework of additive schemes is
proposed and studied by Kelly in \cite{Kelly1985StochModels}, where
the estimate $\hat{N}_{t}$ is updated by the recursion
\begin{equation} \label{eq:est_framework}
\hat{N}_{t+1} = \max\{1, \hat{N}_t + a_0 I(Z_t = 0) + a_1 I(Z_t = 1)
+ a_cI(Z_t = c)\},
\end{equation}
where $a_0$, $a_1$, and $a_c$ are constants and $I(A)$ is the
indicator function of event $A$.

With the estimation, the BS sets the transmission probability to
$p_t = 1/\hat{N}_t$ for all backlogged devices, and thus the offered
load $\rho = N_t p_t = N_t/\hat{N}_t$, representing the average
number of devices attempting to access the channel. To stabilize and
optimize the S-ALOHA system, $\hat{N}_t$ needs to drift towards the
actual number of backlogged devices $N_t$, especially when $N_t$ is
large. When $N_t = n$ and $\hat{N}_t = \hat{n}$, the drift of the
estimate can be calculated as follows \cite{Kelly1985StochModels}:
\begin{align} \label{eq:est_drift}
&\quad E[\hat{N}_{t+1}-\hat{N}_t|N_t = n, \hat{N}_t=
\hat{n}]~~~~~~~~~~~~\nonumber\\
&=(a_0-a_c)\left(1-\frac{1}{\hat{n}}\right)^n +
(a_1-a_c)\frac{n}{\hat{n}}\left(1-\frac{1}{\hat{n}}\right)^{n-1}
+a_c \nonumber\\
& \to (a_0-a_c)e^{-\rho} + (a_1-a_c)\rho e^{-\rho} + a_c \overset
{\rm def}{=} \phi(\rho),\nonumber
\end{align}
as $n \to \infty$, with $n/\hat{n} = \rho$ fixed.

By properly choosing the parameters $a_i$ ($i = 0, 1, c$) such that $\phi(\rho) < 0$ if $\rho <1$ and
$\phi(\rho) > 0$ if $\rho >1$, the estimate $\hat{N}_t$ will drift towards the true value and thus the S-ALOHA system can be stabilized. However, these fixed step size schemes are not suitable for systems with bursty traffic. When the estimate $\hat{N}_t$ deviates far away from the true value $N_t$, we have $\lim_{\rho \to 0} \phi(\rho) = a_0$ and $\lim_{\rho \to \infty} \phi(\rho) = a_c$. These limits indicate that the drift tends to be a constant even when the deviation is large, which could result in a large tracking time. Thus, it is necessary to design fast estimation schemes for event-driven M2M communication.

\subsection{Framework of FASA}
As analyzed in the previous subsection, fixed step size estimation
schemes such as PB-ALOHA may not be able to adapt in a timely manner
for systems with bursty traffic because it always uses a constant
step size even when the estimate is far away from the true value. We
note that in addition to the access result in the previous slot, the
access results in several consecutive slots will be helpful for
improving the estimation as they could reveal additional information
about the true value. Intuitively, collisions in several consecutive
slots are likely caused by a significant underestimation, i.e.,
$\hat{N}_t \ll N_t$, and the BS should aggressively increase its
estimate. In contrast, several consecutive idle slots may indicate
that the estimate $\hat{N}_t \gg N_t$, and it should be reduced
aggressively.

Motivated by this intuition, we propose a FASA scheme that updates $\hat{N}_t$ as follows:
\begin{equation} \label{eq:fasa_scheme} \hat{N}_{t+1} =
\begin{cases}
\max\{1, \hat{N}_{t} -1 - h_0(\nu)(K_{0,t} \wedge  k_{m})^{\nu}\},&{\rm if}~Z_t=0\\
\hat{N}_{t},&{\rm if}~Z_t=1\\
\hat{N}_{t} + \frac{1}{e-2}  + h_c(\nu)(K_{c,t}\wedge
k_{m})^{\nu},&{\rm if}~Z_t=c
\end{cases}
\end{equation}
where $K_{0,t}$ and $K_{c,t}$ are the numbers of consecutive idle
and collision slots up to slot $t$, respectively; $k_m > 1$ is an
integer and  $K \wedge k_{m} = \min\{K, k_{m}\}$; $\nu > 0$ is the
parameter that controls the adjusting speed; $h_0(\nu)$ and
$h_c(\nu)$ are functions of $\nu$ that guarantee the right direction
of the estimation drift. In order to make the scheme implementable
and its stability analysis tractable, we bound the update step size
with $k_m$ in this paper, which is different from that we proposed
in \cite{Wu2011FASA}. However, the two schemes are almost the same
as long as we choose a sufficiently large $k_m$.

\subsection{Design of $h_0(\nu)$ and $h_c(\nu)$}

The functions $h_0(\nu)$ and $h_c(\nu)$ are crucial for guaranteeing the convergence of the FASA scheme. Next we design $h_0(\nu)$ and $h_c(\nu)$ by analyzing the drift of
estimate $\hat{N}_t$. According to the structure of the proposed
scheme, the evolution of $\hat{N}_t$ depends not only on the access
result in the previous slot, but also results in the past $k_m$
slots. Therefore, unlike the fixed step size schemes, accurate drift
analysis is impractical for FASA because of its memory property.
Thus, in order to make the problem tractable, we resort to
approximation based on Lemma \ref{thm:prob_approx}, which indicates
the feasibility for approximating the distribution of access results
in the past $k_m$ slots with the network status in the previous slot.

\begin{lemma} \label{thm:prob_approx}
For given $\epsilon > 0$, $\Delta n$, and $\Delta \hat{n}$, there
exists some $M > 0$, such that for any $(n, \hat{n}) \in H_M$, where
$H_M = \{(n, \hat{n}): n \geq M~{\rm or}~ \hat{n} \geq M, n + \Delta
n \geq 0, \hat{n} + \Delta{\hat{n}} \geq 1\}$, we have
\begin{align}
&\left|e^{-\rho} - (1 - \frac{1}{\hat{n} + \Delta \hat{n}})^{n +
\Delta n}\right| \leq \epsilon, \label{eq:idle_prob_gapbound}\\
&\left|\rho e^{-\rho} - \frac{n+\Delta n}{\hat{n} + \Delta
\hat{n}}(1 - \frac{1}{\hat{n} + \Delta \hat{n}})^{n + \Delta n -
1}\right| \leq \epsilon,\label{eq:succ_prob_gapbound}
\end{align}
where $\rho = n/\hat{n}$.
\end{lemma}

\proof See Appendix \ref{app:proof_of_prob_approx}.

It is noticed that the distribution of the access result $Z_s$ in
slot $s$ is decided by $N_s$ and $\hat{N}_s$. In addition, for any
given $s \in \{t-k_m, t-k_m+1, \ldots, t-1\}$, we have $|\hat{N}_s
-\hat{N}_t| \leq k_m /(e-2) + k_m^{1+\nu}
\max\{h_0(\nu),h_c(\nu)\}$, and $|N_s - N_t|$ is bounded with high
probability, i.e., ${\rm Pr}(|N_s - N_t| \leq B) \to 1$ as $B \to
\infty$. Thus, according to Lemma \ref{thm:prob_approx}, when $N_t$
and $\hat{N}_t$ are known and at least one of them is sufficiently
large, the distribution of access results in the past $k_m$ slots
can be evaluated approximately, as well as the statistical
characteristics of $K_{0,t}$ and $K_{c,t}$. Therefore, in this
section, we do not assume any knowledge about $K_{0, t}$ or
$K_{c,t}$ in slot $t$ and will approximately calculate the drift of
$\hat{N}_t$ conditioned on $N_t$ and $\hat{N}_t$. In addition, we
assume that $\hat{N}_t$ is large enough in the past $k_m$ slots, and
hence we can approximate $\max\{1,x\}$ in \eqref{eq:fasa_scheme} as
$x$ in the analysis later. Rigorous analysis provided in Section
\ref{sec:stability} will show that the design with these
approximations stabilizes the proposed scheme.

Suppose that in slot $t$, the number of backlogged devices and its
estimate are $N_t = n$, $\hat{N}_t = \hat{n}$, respectively, and
thus the offered load $\rho = n/\hat{n}$. When $n$ or $\hat{n}$ is
large, the drift of estimate $\hat{N}_t$ can be approximated as
\begin{align} \label{eq:est_drift_fasa}
&\quad E[\hat{N}_{t+1}-\hat{N}_t|N_t = n, \hat{N}_t = \hat{n}]\nonumber\\
&\approx \sum_{i \in\{0, 1, c\}} q_i(\rho)E[\Delta\hat{N}_t|(i,
n,\hat{n})]
\end{align}
where $q_0(\rho) = e^{-\rho}$, $q_1(\rho) = \rho e^{-\rho}$, and
$q_c(\rho) = 1 - q_0(\rho) - q_1(\rho)$ are the probabilities of an
{\it{idle}}, {\it{success}}, and {\it{collision}} slot,
respectively; $E[\Delta\hat{N}_t|(i, n,\hat{n})]$ ($i = 0,1,c$) are
the changes in $\hat{N}_t$ resulting from the corresponding updates.

Obviously, $E[\Delta\hat{N}_t|(1, n,\hat{n})] = 0$ since the
estimated number remains unchanged when a packet is successfully
transmitted in slot $t$. On the other hand, without memory about the
access results in the past slots, $K_{0,t}$ and $K_{c,t}$ are
treated as random variables. Hence, $E[\Delta\hat{N}_t|(0,
n,\hat{n})]$ and $E[\Delta\hat{N}_t|(c, n,\hat{n})]$ can be obtained
based on the approximate distributions of $K_{0,t}$ and $K_{c,t}$.

First, to calculate the drift of estimate in an idle slot
$E[\Delta\hat{N}_t|(0, n,\hat{n})]$, suppose that no packet is
transmitted in slot $t$, then the estimated number will be reduced
by $1 + h_0(\nu)(K_{0,t}\wedge k_m)^\nu$. Therefore,
\begin{align} \label{eq:drift_of_est_idle}
&\quad E[\Delta\hat{N}_t|(0, n,\hat{n})] \nonumber\\
&=  - \sum_{k_0 = 1}^{k_m-1}[1 + h_0(\nu) k_0^\nu]{\rm Pr}[K_{0,t} =
k_0|(0, n,\hat{n})]\nonumber\\
&\quad - [1 + h_0(\nu) k_m^\nu]{\rm Pr}[K_{0,t} \geq k_m |(0,
n,\hat{n})].
\end{align}

Notice that $K_{0,t} = k_0 (1 \leq k_0 < k_m)$ holds when slots $t -
k_0 + 1$, $t - k_0 + 2$, \ldots, $t-1$ are all idle while slot $t
-k_0$ is not. Thus, for $1 \leq k_0 < k_m$, we have
\begin{align} \label{eq:consec_idle_k0}
&\quad {\rm Pr}[K_{0,t} = k_0 |(0, n,\hat{n})] \nonumber\\
&= {\rm Pr}[Z_{t - k_0} \neq 0 | (Z_{t-k_0+1}, \ldots ,Z_t,
N_t,N'_t) =
(0, \ldots, 0, n,\hat{n})]\nonumber\\
& \cdot \prod_{s = t-k_0 + 1}^{t - 1} {\rm Pr}[Z_{s} = 0 | (Z_{s+1},
\ldots , Z_t,N_t,N'_t) = (0, \ldots, 0, n,\hat{n})].\nonumber
\end{align}

According to Lemma \ref{thm:prob_approx}, when $N_t = n$ or
$\hat{N}_t = \hat{n}$ are sufficiently large, the distribution of
access results in the past $k_m$ slots can be approximated as that
in the previous slot, i.e.,
\begin{align*} \label{eq:}
&\quad {\rm Pr}[Z_{t - k_0} \neq 0 | (Z_{t-k_0+1}, \ldots ,Z_t,
N_t,N'_t) = (0,
\ldots, 0, n,\hat{n})]\\
& \approx 1 - q_0(\rho),\\
&\quad {\rm Pr}[Z_{s} = 0 | (Z_{s+1}, \ldots , Z_t,N_t,N'_t) = (0,
\ldots, 0,
n,\hat{n})]\\
&\approx  q_0(\rho),~~~s = t - k_0 + 1, t - k_0 + 2, \ldots, t -
1.\nonumber
\end{align*}

Consequently, for $1 \leq k_0 < k_m$,
\begin{equation} \label{eq:consec_idle_k0}
{\rm Pr}[K_{0,t} = k_0 |(0, n,\hat{n})] \approx
q_0^{k_0-1}(\rho)[1-q_0(\rho)].
\end{equation}

Similarly, $K_{0,t} \geq k_m$ holds if slots $t - k_m +1, t-k_m + 1,
t-1$ are all idle and we can approximate the probability as
\begin{equation} \label{eq:consec_idle_km}
{\rm Pr}(K_{0,t} \geq k_m) \approx q_0^{k_m-1}(\rho).
\end{equation}

Substituting (\ref{eq:consec_idle_k0}) and (\ref{eq:consec_idle_km})
into (\ref{eq:drift_of_est_idle}), we can calculate the drift of
$\hat{N}_t$ in an idle slot as follows:
\begin{align} \label{eq:drift_of_est_idle_2}
E[\Delta\hat{N}_t|(0, n,\hat{n})] &\approx  -\sum_{k_0 =
1}^{k_m-1}[1+h_0(\nu)k_0^\nu]q_0^{k_0-1}(\rho)[1-q_0(\rho)]
\nonumber\\
&\quad - [1+h_0(\nu)k_m^\nu]q_0^{k_m-1}(\rho)\nonumber\\
&=-[1+h_0(\nu)\mu(\nu, q_0(\rho),k_{m})],
\end{align}
where $\mu(\nu, q,k_{m})$ is defined as
\begin{align} \label{eq:nu_order_moment}
\mu(\nu, q,k_{m}) = \sum_{k = 1}^{k_{m}-1}k^\nu q^{k-1} (1-q) +
(k_{m})^\nu q^{k_{m}-1},
\end{align}
and $\mu(\nu, q_0(\rho),k_{m})$ is the approximate expectation of
$(K_{0,t}\wedge k_m)^\nu$ conditioned on $(Z_t, N_t, \hat{N}_t) =
(0, n, \hat{n})$.

Second, we can calculate the drift of estimate in a collision slot
in a similar fashion as follows:
\begin{align} \label{eq:drift_of_est_coll_2}
&\quad E[\Delta\hat{N}_t|(c, n, \hat{n})] \nonumber\\
&\approx  (e-2)^{-1}+ h_c(\nu)E[(K_{c,t}\wedge k_m)^\nu |(c, n,
\hat{n})]
\nonumber\\
&=(e-2)^{-1}+h_c(\nu)\mu(\nu, q_c(\rho),k_{m}).
\end{align}

Therefore, the drift of estimate for FASA can be approximated by
substituting the expressions of $E[\Delta \hat{N}_t|(i, n,
\hat{n})]$ ($i = 0, 1, c$) into (\ref{eq:est_drift_fasa}):
\begin{align} \label{eq:est_drift_fasa_2}
&\quad E[\hat{N}_{t+1}-\hat{N}_t|N_t = n, \hat{N}_t = \hat{n}]\nonumber\\
&\approx  -q_0(\rho)[1+h_0(\nu)\mu(\nu, q_0(\rho),k_{m})]\nonumber\\
&\quad +q_c(\rho)[(e-2)^{-1}+h_c(\nu)\mu(\nu, q_c(\rho),k_{m})]\nonumber\\
&\overset {\rm def}{=}\varphi(\rho).
\end{align}

In order to keep the offered load $\rho$ staying in the neighborhood
of the optimal value $\rho^* = 1$, it is reasonable to require that
$\varphi(1) = 0$. In other words, letting $q_0^* = q_0(1) = e^{-1}$
and $q_c^* = q_c(1) = 1-2e^{-1}$, we expect that
\begin{align} \label{eq:est_drift_fasa_equilium}
\varphi(1)&= -q_0^*[1 + h_0(\nu)\mu(\nu, q_0^*, k_m)] \nonumber\\
&\quad + q_c^*[(e-2)^{-1} + h_c(\nu)\mu(\nu, q_c^*, k_m)]\nonumber\\
&= -h_0(\nu)q_0^*\mu(\nu, q_0^*,k_m)+ h_c(\nu)q_c^*\mu(\nu,
 q_c^*,k_m)  = 0.
\end{align}

Hence, for given $k_m > 1$, to satisfy the condition in
(\ref{eq:est_drift_fasa_equilium}), we can select the following
$h_0(\nu)$ and $h_c(\nu)$:
\begin{equation} \label{eq:norm_function_idle}
h_0(\nu) = \eta [q_0^*\mu(\nu, q_0^*,k_m)]^{-1},
\end{equation}
\begin{equation} \label{eq:norm_function_coll}
h_c(\nu) = \eta [q_c^*\mu(\nu, q_c^*,k_m)]^{-1},
\end{equation}
where $\eta > 0$ is a constant.

The chosen $h_0(\nu)$ and $h_c(\nu)$ guarantee that $\varphi(1)= 0$
and thus provide a necessary condition for FASA to track the number
of backlogged devices. Furthermore, Theorem
\ref{thm:equilibrium_of_fasa} shows a desirable property of FASA,
with which the estimated number $\hat{N}_t$ roughly drifts towards
to the true value $N_t$, eventually yielding $\rho = N_t/\hat{N}_t
\approx 1$.

\begin{theorem}\label{thm:equilibrium_of_fasa}
Given that $h_0(\nu)$ and $h_c(\nu)$ are defined by
(\ref{eq:norm_function_idle}) and (\ref{eq:norm_function_coll}),
respectively, the approximate drift of FASA $\varphi(\rho)$ is a
strictly increasing function of $\rho$. In addition, $\varphi(\rho)
< 0$ when $0 < \rho <1$ and $\varphi(\rho)>0$ when $\rho
>1$.
\end{theorem}

\proof See Appendix \ref{app:proof_of_equilibrium_of_fasa}.

\begin{figure}[htbp]
\begin{center}
\includegraphics[angle = 0,width = 0.8\linewidth]{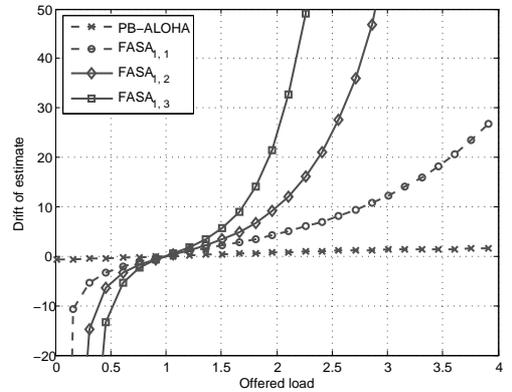}
\caption{Drift of estimation.} \label{fig:drift_of_est}
\end{center}
\vspace{-0.5cm}
\end{figure}

In order to understand better the behavior of the scheme, we now present the approximate drift of estimate for FASA with $\eta
= 1$ and $\nu = 1,2,3$. Assume that $k_m$ is large, and thus the
distribution of $K_{i,t}$ $(i = 0, c)$ can be approximated as a
geometrical distribution with success probability $1-q_i(\rho)$. In
addition, for $\nu \in \mathbb{Z}_+$, $\mu(\nu, q_i(\rho), k_m)$ is
approximately the $\nu$th-moment of a geometrically distributed
random variable with success probability $1-q_i(\rho)$, and its
closed-form expression can be obtained. Consequently, the results
obtained in our previous work \cite{Wu2011FASA} can be applied
directly. Fig.~\ref{fig:drift_of_est} shows the approximate drift of
estimation versus offered load, where the subscript of FASA
represents value of $(\eta, \nu)$. It can be observed from the
figure that when the estimated number deviates far away from the
actual number of backlogged devices, i.e., $\rho \approx 0$ or $\rho
\gg 1$, FASA adjusts its step size accordingly, while PB-ALOHA still
uses the same step size. Therefore, using FASA results in much
shorter adjusting time than PB-ALOHA, and thus could improves the
performance of M2M communication systems with bursty traffic. Note that the drifts of multiplicative schemes such as $Q^+$-Algorithm are not illustrated here since they depend on not only the offered load $\rho$ but also the estimate $\hat{N}_t$.

\section{Stability Analysis of FASA}\label{sec:stability}
In this section, we use drift analysis to study the stability of the
proposed FASA scheme. The M2M traffic is modeled as an interrupted
Poisson process presented in Section \ref{sec:sys_model}. In fact,
as we will discuss later, the proposed scheme can be stable under
other more general arrival processes.

Unlike traditional adaptive schemes, the access results in the past
consecutive slots are used in FASA to accelerate the speed of
tracking, which makes it difficult to obtain the accurate drift of
estimate. However, from the stability point of view, we concern
mostly the scenarios where the number of backlogged devices or its
estimate is large and hence approximation can be applied in these
cases. To deal with the issues caused by the memory property of
FASA, we analyze its $T$-slot drifts rather than the 1-slot drifts,
which are introduced for analyzing traditional adaptive ALOHA
schemes
\cite{Hajek1982ITAC,Hajek1982DriftAnal,Tsitsiklis1987StabAnal,Gurcan2001ITVT}.
By constructing a virtual sequence, we show that the $T$-slot drifts
of FASA have the properties required for stabilizing the system as
long as the number of backlogged or its estimate is sufficiently
large, and these are similar to the properties of PB-ALOHA.
Therefore, with slight modification, the Lyapunov function based
method proposed for PB-ALOHA \cite{Tsitsiklis1987StabAnal} can be
used to prove the stability of FASA.


Consider the FASA scheme proposed in \eqref{eq:fasa_scheme} under
interrupted Poisson arrival process with average arrival rate
$\bar{\lambda}$. We define a sequence $X_t = (N_t, \hat{N}_t, K_t)$,
where $K_t$ represents the memory of access results in the past
consecutive slots and is defined as $K_0 = 0$ and for $t> 0$,
\begin{equation*} \label{eq:} K_t=
\begin{cases}
- (K_{0,t-1} \wedge  k_{m}),&{\rm if}~Z_{t-1}=0,\\
0,&{\rm if}~Z_{t-1}=1,\\
K_{c,t-1}\wedge k_{m},&{\rm if}~Z_{t-1}=c.
\end{cases}
\end{equation*}
Recall that $K_{0,t-1}$ and $K_{c,t-1}$ is the number of consecutive
idle and collision slots up to slot $t-1$. Given initial value $X_0
= (0, 1, 0)$, each component of $X_t$ evolves as follows when $t >
0$ :
\begin{align}
&K_{t+1} =
\begin{cases}
-(|K_{t}|+1)\wedge  k_m,&{\rm if}~K_{t} < 0,Z_{t}=0,\\
-1,&{\rm if}~K_{t} \geq 0,Z_{t}=0,\\
0,&{\rm if}~Z_{t}=1,\\
1, &{\rm if}~K_{t} \leq 0,Z_{t}=c,\\
(K_{t} + 1)\wedge k_m,&{\rm if}~K_{t} > 0,Z_{t}=c,\\
\end{cases}\label{eq:evol_k}\\
&N_{t+1} = \max\{0, N_{t} + A_{t} - I(Z_{t} =
1)\}\label{eq:evol_n}\\
&\hat{N}_{t+1} =
\begin{cases}
\max\{1, \hat{N}_{t+1} -1 - h_0(\nu)|K_{t+1}|^\nu\},&{\rm if}~Z_{t}=0,\\
\hat{N}_{t},&{\rm if}~Z_{t}=1,\\
\hat{N}_{t} + \frac{1}{e-2}  + h_c(\nu)(K_{t+1})^\nu, &{\rm
if}~Z_{t}=c.
\end{cases}\label{eq:evol_nhat}
\end{align}

It is easy to verify that $X_t = (N_t, \hat{N}_t, K_t)$ is a Markov
chain on a countable state space $\mathbb{S}_X$. The main result of
this section reveals the geometrical ergodicity
\cite{Hajek1982DriftAnal} of $X_t$, which is described in Theorem
\ref{thm:stability_of_fasa}. As pointed out in \cite{Hajek1982DriftAnal}, the geometrical ergodicity is a weaker form of ergodicity and indicates the existence of steady distribution for each initial state.

\begin{theorem}\label{thm:stability_of_fasa}
If $0 < \bar{\lambda} < e^{-1}$, $k_m > 1$, $h_0(\nu)$ and
$h_c(\nu)$ are given by (\ref{eq:norm_function_idle}) and
(\ref{eq:norm_function_coll}), then the Markov Chain $X_t$ is
geometrically ergodic.
\end{theorem}

\begin{proof}
The proof of Theorem \ref{thm:stability_of_fasa} is based on the
drift analysis. Specifically, the proof involves three steps, which
are outlined as follows and presented afterwards:

{\it Step 1 - Approximation of drifts:} To deal with the impact of
the memory in the proposed scheme, rather than 1-slot drifts in the
existing works, we study the $T$-slot drifts for our scheme, which
are then approximated by constructing a virtual sequence $X'_{t+s}$
conditioned on the state of $X_t$ in slot $t$.

{\it Step 2 - Property analysis of drifts:} Based on the
approximation of $T$-slot drifts for FASA, we obtain the properties
of the $T$-slot drifts required for guaranteeing the stability of
the scheme.

{\it Step 3 - Stability analysis based on Lyapunov function:} The
Lyapunov function defined in \cite{Tsitsiklis1987StabAnal} is
adopted for the proposed scheme. Then with the the properties
obtained in Step 2, we show the geometrical ergodicity of $X_t$ by
analyzing the drifts of the Lyapunov function.

\subsection*{Step 1 - Approximation of Drifts}
In order to analyze the stability of FASA, we evaluate the change of
$X_t$ from slot $t$ to slot $t +T$. Let $\tilde{N}_t = \hat{N}_t -
N_t$ denote the estimate error in slot $t$. Conditioned on the state
of $X_t$, we define the $T$-slot drifts of $N_t$, $\hat{N}_t$, and
$\tilde{N}_t$ as follows:
\begin{align}
d_T(n, \hat{n}, k)&= E[N_{t+T} - N_t | X_t = (n, \hat{n},k)],\label{eq:T_slot_drifts_fasa_n}\\
\hat{d}_T(n, \hat{n}, k) &= E[\hat{N}_{t+T} - \hat{N}_t | X_t = (n,
\hat{n},k)],\label{eq:T_slot_drifts_fasa_nhat}\\
\tilde{d}_T(n, \hat{n}, k) &= E[\tilde{N}_{t+T} - \tilde{N}_t | X_t
= (n, \hat{n},k)] \nonumber\\
& =\hat{d}_T(n, \hat{n}, k) - d_T(n, \hat{n},
k).\label{eq:T_slot_drifts_fasa_gap}
\end{align}
%

It is difficult to calculate the drifts defined above and we try to
obtain the properties of them by introducing an approximate version
of $X_t$. Let $\{Z'_{t+s}: s \in \mathbb{Z}_+\}$ be a ternary
independently and identically distributed (i.i.d.) random sequence,
whose distribution is given by
\begin{equation*}
{\rm Pr}(Z'_{t+s} = i) = q_i(\rho), ~~i = 0,1,c,
\end{equation*}
where $\rho = n/\hat{n}$. Then we construct a virtual sequence
$X'_{t+s} = (N'_{t+s},\hat{N}'_{t+s}, K'_{t+s})$ based on $X_t$ and
$Z'_{t+s}$ as follows: when $s = 0$, $X'_t = X_t = (n,\hat{n},k)$;
when $s
> 0$, $K'_{t+s}$, $N'_{t+s}$, and $\hat{N}'_{t+s}$ are updated in a
similar way as (\ref{eq:evol_k}) - (\ref{eq:evol_nhat}),
respectively, with $Z_t$ replaced with $Z'_{t+s}$. However, unlike
the updates in $X_t$, we allow $N'_{t+s}$ to be negative and
$\hat{N}'_{t+s}$ to be less than 1.

Obviously, $X'_{t + s}$ is a Markov chain and its transition
probabilities are fixed and determined by the state of $X_t$. We
define its $T$-slot drifts $d'_T(n, \hat{n}, k)$, $\hat{d}'_T(n,
\hat{n}, k)$, and $\tilde{d}'_T(n, \hat{n}, k)$ similarly to
\eqref{eq:T_slot_drifts_fasa_n} - \eqref{eq:T_slot_drifts_fasa_gap}.
When $T$ is given, we show in Lemma \ref{thm:approx_drift} that the
drifts of $X'_{t+s}$ can be used to approximate the drifts of $X_t$
from slot $t$ to slot $t+T$, when either $N_t$ or $\hat{N}_t$ is
sufficiently large.


\begin{lemma}\label{thm:approx_drift}
Given $T > 0$ and $\epsilon > 0$, there exists some $M > 0$, such
that if $N_t = n \geq M$ or $\hat{N}_t = \hat{n} \geq M$, then
\begin{align}
|d_T(n, \hat{n}, k) - d'_T(n, \hat{n}, k)|\leq \epsilon,\label{eq:approx_drift_gap_n}\\
|\hat{d}_T(n, \hat{n}, k) - \hat{d}'_T(n, \hat{n}, k)|\leq
\epsilon.\label{eq:approx_drift_gap_nhat}
\end{align}
\end{lemma}

\proof See Appendix \ref{app:proof_of_approx_drift}.

According to Lemma \ref{thm:approx_drift},  for given $T$, the
differences between the $T$-slot drifts of $X'_{t + s}$ and $X_{t + s}$ can be
made as close to zero as desired by letting $n$ or $\hat{n}$ be
sufficiently large. Next, we evaluate the drifts
of $X'_{t+s}$ for obtaining the properties of drifts for FASA in Step 2.\\

\subsubsection{$d'_T(n, \hat{n}, k)$}

Since in the virtual sequence $X'_{t+s}$, $N'_{t+s}$ is allowed to
be negative, we can easily have
\begin{align*} \label{eq:}
d'_T(n, \hat{n}, k)& = E[\sum_{s = 0}^{T-1} (A_{t+s}-Z'_{t+s})|X'_t
= (n, \hat{n},k)]\\
&= T(\bar{\lambda} - \rho e^{-\rho}).
\end{align*}


\subsubsection{$\hat{d}'_T(n, \hat{n}, k)$}

In slot $t+s$, the update of $\hat{N}'_{t+s}$ depends on both $K'_{t
+s}$ and $Z'_{t +s}$. Notice that the sequence $K'_{t +s}$ is a
Markov chain on a finite state space $\mathbb{S}_K = \{-k_m, -k_m +
1, \ldots, 0, \dots, k_m\}$. Since the distribution of $Z'_{t +s}$
is fixed, by showing the ergodicity of $K'_{t +s}$, we are able to
approximate the $T$-slot drift by analyzing the stationary behavior
of $K'_{t +s}$. Specifically, the transition of $K'_{t +s}$ depends
on the value of $Z'_{t +s}$ and the 1-step transition probabilities
is given by
\begin{equation*}\label{eq:}
p_{k j}(\rho)= \begin{cases}
q_0(\rho),&{\rm if}~k \leq 0, j = \max\{k - 1, -k_m\},\\
q_0(\rho),&{\rm if}~k > 0, j = -1,\\
q_1(\rho),&{\rm if}~j = 0,\\
q_c(\rho),&{\rm if}~k < 0, j = 1,\\
q_c(\rho),&{\rm if}~k \geq 0, j = \min\{k + 1, k_m\},\\
0,&{\rm else.}
\end{cases}
\end{equation*}

It is easy to verify that when $\rho > 0$, $K'_{t+s}$ is irreducible
and aperiodic. Thus, $K'_{t+s}$ is ergodic and there is a unique
stationary distribution. Now we study the stationary distribution of
$K'_{t+s}$ and the drift of $\hat{N}'_{t+s}$ in the steady state.
Define a $1\times (2k_m +1)$ vector as follows:
\begin{equation*}\label{eq:}
\pmb{\pi}(\rho) = [\pi_{-k_m}(\rho), \pi_{-k_m + 1}(\rho), \ldots,
\pi_0(\rho), \ldots, \pi_{k_m}(\rho)]\nonumber,
\end{equation*}
where the elements are given by
\begin{equation*}\label{eq:}
\pi_k(\rho) = \begin{cases}
q_0^{k_m}(\rho),&{\rm if}~k = -k_m,\\
q_0^{|k|}(\rho)[1-q_0(\rho)],&{\rm if}~-k_m + 1 \leq k \leq -1,\\
q_1(\rho),&{\rm if}~k = 0,\\
q_c^{k}(\rho)[1-q_c(\rho)],&{\rm if}~1 \leq k \leq k_m-1,\\
q_c^{k_m}(\rho),&{\rm if}~k = k_m.
\end{cases}
\end{equation*}
It can be verified that $\sum_{j = -k_m}^{k_m}\pi_j (\rho) = 1$ and
$\pi_j (\rho)= \sum_{k = -k_m}^{k_m} \pi_k (\rho) p_{kj}(\rho)$ for
all $k \in \mathbb{S}_K$. Hence, $\pmb{\pi}(\rho)$ is the stationary
distribution of $K'_{t + s}$. Using the expression of
$\pmb{\pi}(\rho)$, we can verify that $\varphi(\rho)$ defined in
(\ref{eq:est_drift_fasa_2}) represents the stationary drift of
$\hat{N}'_{t+s}$, which is the 1-slot drift of $\hat{N}'_{t+s}$ when
$K'_{t + s}$ is in the steady state. Consequently, with the
ergodicity of $K'_{t+s}$, we have
\begin{equation*}\label{eq:}
\lim_{T \to \infty}  \frac{1}{T} \hat{d}'_T(n, \hat{n}, k)=
\varphi(\rho).\nonumber
\end{equation*}

Moreover, in order to use Lemma \ref{thm:approx_drift}, we expect to
find a common $T$  such that \eqref{eq:approx_drift_gap_n} and
\eqref{eq:approx_drift_gap_nhat} hold for some given $\epsilon$ and
for all $(n, \hat{n}, k) \in \mathbb{S}_X$, which requires the
uniform convergence of $\frac{1}{T} \hat{d}'_T(n, \hat{n}, k)$. In
fact, by analyzing the evolution of $K'_{t+s}$, we can show that
$\frac{1}{T} \hat{d}'_T(n, \hat{n}, k)$ converges uniformly in
$(n,\hat{n},k)$. First, by multiplying the transition probability
matrix $k_m$ times or analyzing the event that $K'_{t+k_m} = j$, we
can see that for any $k, j \in \mathbb{S}_K$, the $k_m$-step
transition probability $p^{(k_m)}_{k,j}(\rho) = \pi_j(\rho)$.  For
example, for any $k \in \mathbb{S}_K$ and $j \in (-k_m, 0)$, $K'_{t
+ k_m} = j$ holds if and only if $Z'_{t + s} = 0$ for all $s =
k_m-1, k_m-2, k_m -j+1$, while $Z'_{t + k_m -j} \neq 0$, so
$p^{(k_m)}_{k,j}(\rho) = q_0^{|j|}(\rho)[1-q_0(\rho)] =
\pi_j(\rho)$. Consequently, for any state $k\in \mathbb{S}_K$, we
have
\begin{equation*}\label{eq:}
{\rm Pr}(K'_{t+s} = j) = \pi_j(\rho),~ \mbox{$s \geq k_m$},\nonumber
\end{equation*}
and thus when $s \geq k_m$ the drift of $\hat{N}'_{t+s}$ in each
slot is exactly $\varphi(\rho)$. Then, with the fact that for any
$(n, \hat{n}, k) \in \mathbb{S}_X$,
\begin{align*}\label{eq:}
&\quad\left|\hat{d}'_{k_m-1}(n, \hat{n}, k)-(k_m -1)\varphi(\rho)\right|\\
&\leq (k_m -1)\left[\frac{1}{e-2} + k_m^\nu \max\{h_0(\nu),
h_c(\nu)\}\right],
\end{align*}
we know that as $T$ tends to infinity, $\frac{1}{T} \hat{d}'_T(n,
\hat{n}, k)$ converges to $\varphi(\rho)$ uniformly in $(n, \hat{n},
k)$. Thus, the difference between $\frac{1}{T} \hat{d}'_T(n,
\hat{n}, k)$ and $\varphi(\rho)$ can be made as close to zero as
desired by choosing a common $T$ for all $(n, \hat{n},k) \in
\mathbb{S}_X$.


%
%
%
%
%
%

\subsubsection{$\tilde{d}'_T(n, \hat{n}, k)$}

Since $\tilde{d}'_T(n, \hat{n}, k) = \hat{d}'_T(n, \hat{n},
k)-d'_T(n, \hat{n}, k)$, we introduce the following function to
approximate $\frac{1}{T}\tilde{d}'_T(n, \hat{n}, k)$:
\begin{align*} \label{eq:}
\psi(\rho, \bar{\lambda})&= \varphi(\rho) - (\bar{\lambda} -
\rho e^{-\rho})\nonumber\\
&=[\frac{1}{e-2} + h_c(\nu)\mu(\nu, q_c(\rho),k_{m})](1- e^{-\rho}-\rho e^{-\rho}) \nonumber\\
& \quad - [1 + h_0(\nu)\mu(\nu, q_0(\rho),k_{m})]e^{-\rho}+\rho
e^{-\rho} - \bar{\lambda}.~~~~\nonumber
\end{align*}

With the uniform convergence of $\frac{1}{T}\hat{d}'_T(n, \hat{n},
k)$, we know that for any given $\bar{\lambda} > 0$, as $T \to
\infty$,
\begin{equation*}\label{eq:}
\frac{1}{T} \tilde{d}'_T(n, \hat{n}, k)\to \psi(\rho,
\bar{\lambda})\nonumber
\end{equation*}
uniformly in $(n, \hat{n}, k)$.
%

\subsection*{Step 2 - Property Analysis of Drifts}
The evolution of estimate error $\tilde{N}_t$ is critical for
showing the stability of the scheme. We first show in Lemma
\ref{thm:prop_est_err_drift} that the approximate drift $\psi(\rho,
\bar{\lambda})$ has the same properties as those for the 1-slot
drift of PB-ALOHA and then present the required properties of
$T$-slot drifts of FASA in Lemma \ref{thm:strictly_non_zero_drift}.

\begin{lemma} \label{thm:prop_est_err_drift} Given $k_{m}$ as a positive integer,
 $\psi(\rho, \bar{\lambda})$ has the following properties:

a) For any $\bar{\lambda}$, the function $\psi(\rho,
\bar{\lambda})$ is strictly increasing in $\rho$.

b) For any $\bar{\lambda} \in (0, e^{-1}]$, there exists a unique
$\rho = \omega(\bar{\lambda}) \in (0,1]$, such that $\psi(\rho,
\bar{\lambda}) = 0$.

c) If $\bar{\lambda} \in (0, e^{-1})$, then $\omega(\bar{\lambda})
e^{-\omega(\bar{\lambda})} > \bar{\lambda}$.
\end{lemma}

\proof See Appendix \ref{app:proof_of_est_err_drift}.

Let $\beta = \omega(\bar{\lambda})$ denote the root of $\psi(\rho,
\bar{\lambda}) = 0$ for given $\bar{\lambda}$. Similarly to the
method in \cite{Tsitsiklis1987StabAnal}, we partition the state
space into the following four parts:
\begin{align}\label{eq:}
S_{\gamma, M}& = \{(n,\hat{n}, k): n\geq M~ {\rm or}~\hat{n}
\geq M,& \nonumber\\
&~~~~~\beta-\gamma \leq \frac{n}{\hat{n}} \leq 1 + \gamma, k \in
\mathbb{S}_K\},&\nonumber \\
R_{\gamma, M}^-& = \{(n,\hat{n},k): \hat{n} \geq M,
\frac{n}{\hat{n}} < \beta-\gamma, k \in
\mathbb{S}_K \},&\nonumber \\
R_{\gamma, M}^+ &= \{(n,\hat{n},k): n \geq M, \frac{n}{\hat{n}}
> 1+ \gamma, k \in
\mathbb{S}_K \},&\nonumber\\
Q_M& = \left\{(n,\hat{n}, k): n < M, \hat{n} < M, k \in \mathbb{S}_K
\right\},&\nonumber
\end{align}
and let $R_{\gamma, M} =R_{\gamma, M}^- \cup R_{\gamma, M}^+$.

With Lemmas \ref{thm:approx_drift} and \ref{thm:prop_est_err_drift},
we present the properties of the $T$-slot drifts in these regions in
the following lemma.

\begin{lemma} \label{thm:strictly_non_zero_drift} There exist some
$\gamma > 0$, $\delta > 0$, $T > 0$, and $M > 0$,  such that
$5\gamma < \beta$ and
\begin{align}
d_T(n, \hat{n}, k)& \leq -T\delta, &\forall (n, \hat{n}, k) &\in S_{5\gamma, M},\label{eq:negative_drift_n}\\
\tilde{d}_T(n, \hat{n}, k)& \leq -T\delta,& \forall (n, \hat{n}, k)
&\in
R_{\gamma, M}^-, \label{eq:negative_drift_n_tilde}\\
\tilde{d}_T(n, \hat{n}, k)& \geq T\delta,& \forall (n, \hat{n}, k)
&\in R_{\gamma, M}^+.\label{eq:positive_drift_n_tilde}
\end{align}
\end{lemma}

\proof See Appendix \ref{app:proof_of_strictly_non_zero_drift}.

Intuitively, according to Lemma \ref{thm:strictly_non_zero_drift}, when the estimate $\hat{N}_t$ is close enough to $N_t$, positive number of devices will access successfully and leave the network in the following $T$ slots. On the other hand, the deviation of the estimate $\hat{N}_t$ from $N_t$ is expected to decrease when it is larger than a certain threshold. These properties guarantee the stability of FASA, as presented in Step 3.

\subsection*{Step 3 - Stability Analysis Based on Lyapunov Function}
Lemma \ref{thm:strictly_non_zero_drift} shows that with a
sufficiently large $T$, the $T$-slot drifts have similar properties
to the drifts of PB-ALOHA. Hence, when observing the system every
$T$ slots, the Lyapunov function based method for PB-ALOHA can be
used for analyzing the stability of FASA. Next, we provide an
outline of using the Lyapunov function based method to prove the
stability of FASA. For more details about this method, it is
recommended to refer to \cite{Tsitsiklis1987StabAnal}.

Assume that $T$, $M$, $\gamma$, and $\delta$ are fixed and that
inequations \eqref{eq:negative_drift_n} -
\eqref{eq:positive_drift_n_tilde} hold. We use the Lyapunov function
defined in \cite{Tsitsiklis1987StabAnal}:
\begin{align} \label{eq:}
V(n,\hat{n},k) &= \max \left\{n,
\frac{1+3\gamma}{3\gamma}(n-\hat{n}), \frac{\beta -
3\gamma}{1-\beta+3\gamma}(\hat{n}-n) \right\}\nonumber \\
&=
\begin{cases} n, &\mbox{if $(n,\hat{n},k) \in S_{3\gamma,M}$},\\
\frac{1+3\gamma}{3\gamma}(n-\hat{n}), &\mbox{if $(n,\hat{n},k)\in R^+_{3\gamma,M}$},\\
\frac{\beta - 3\gamma}{1-\beta + 3\gamma}(\hat{n}-n), &\mbox{if
$(n,\hat{n},k) \in R^-_{3\gamma,M}$}.
\end{cases}
\end{align}
We will show that if $J$ is sufficiently large, there exists some
$\Delta > 0$ such that
\begin{align} \label{eq:negative_drift_Lyapunov}
&E[V(N_{t+JT}, \hat{N}_{t + JT}, K_{t + JT}) - V(N_{t}, \hat{N}_{t},
K_{t})+ \Delta;\nonumber\\
&~~~(N_t, \hat{N}_t, K_t) \notin Q_{M+(JT)^2}|\mathcal {F}_t] \leq
0,
\end{align}
where $\mathcal {F}_t$ is the $\sigma$-field generated by
$\{A_{s-1}, N_s, \hat{N}_s, K_s: s\leq t\}$ and for random variable
$X$ and event $A$, the notation $E[X;A|\mathcal {F}]$ stands for
$E[XI(A)|\mathcal {F}]$.

For given $t \geq 0$ and integer $J$, let
\begin{equation*}\label{eq:}
\tau_{J} = \min \{j \geq 0: \sum_{s = 0}^{jT}A_{t+s} \geq
JT\}.\nonumber
\end{equation*}
Similarly to \cite{Tsitsiklis1987StabAnal}, we then analyze the
drift of the Lyapunov function by considering the unlikely event
$\{\tau_J \leq J\}$ and likely event $\{\tau_J > J\}$ separately.

Using Chernoff bound \cite{Mitzenmacher2005ProbCom}, we can show
that the following results also hold for interrupted Poisson
process:
\begin{equation}\label{eq:arr_process_exp_decay_1}
\lim_{J\to\infty}{\rm Pr}(\tau_J \leq J) = 0,
\end{equation}
\begin{equation}\label{eq:arr_process_exp_decay_2}
\lim_{J\to\infty}E\left[l_1 JT + l_2 \sum_{s = 0}^{JT}A_{t+s};
\tau_J \leq J\right] = 0,
\end{equation}
where $l_1$, $l_2$ are arbitrary given constants. Thus, for any $(n,
\hat{n},k)\in \mathbb{S}_X$, as $J \to \infty$, we have
\begin{align} \label{eq:drif_of_Lyapunov_unlikely}
&E[|V(N_{t+JT}, \hat{N}_{t + JT}, K_{t + JT}) -
V(N_{t}, \hat{N}_{t}, K_{t})|; \nonumber\\
&\qquad \qquad \tau_J \leq J | X_t = (n, \hat{n},k)] \to 0,
\end{align}
implying that this expectation can be made as close to 0 as desired
by choosing a sufficiently large $J$.

Now consider the event $\tau_J > J$. Based on the value of $X_t =
(n, \hat{n},k)$, we study the drift of the Lyapunov function in the
following five cases:

{\it a)} $(n,\hat{n}, k)\in S_{2\gamma, M+(JT)^2}$;

{\it b)} $(n,\hat{n},k)\in R_{4\gamma, M+(JT)^2}^+$;

{\it c)} $(n, \hat{n},k)\in R_{4\gamma, M+(JT)^2}^-$;

{\it d)} $(n,\hat{n}, k)\in R_{4\gamma, M+(JT)^2} \cap R_{2\gamma,
M+(JT)^2}^+$;

{\it f)} $(n,\hat{n}, k)\in R_{4\gamma, M+(JT)^2} \cap R_{2\gamma,
M+(JT)^2}^-$.

In any of these cases, following the approach in
\cite{Tsitsiklis1987StabAnal}, we can show that when $J$ is
sufficiently large, there exists some $\Delta
> 0$, such that inequation \eqref{eq:negative_drift_Lyapunov} holds.

Take case a) as an example. According to Lemma 3.4 in
\cite{Tsitsiklis1987StabAnal}, if $X_t = (n, \hat{n},k) \in
S_{2\gamma, M+(JT)^2}$, then we choose a sufficiently large $J$,
such that $(N_{t+s},\hat{N}_{t+s}, K_{t+s})\in S_{3\gamma, M}$ for
all $s = 0,1,\ldots,JT$, and
\begin{equation*}\label{eq:}
V(N_{t+jT}, \hat{N}_{t+jT}, K_{t+jT}) = N_{t+jT},~~\mbox{for all
$j\in[0,J]$}.
\end{equation*}
Thus, choosing a sufficiently large $J$ such that ${\rm Pr}(\tau_J >
J)
> 1/2$, we have
\begin{align} \label{eq:drift_of_Lyapunov_likely_case1}
&\quad E[V(N_{t+JT}, \hat{N}_{t+JT}, K_{t+JT}) - V(N_{t},
\hat{N}_{t},
K_{t}); \nonumber\\
&\qquad \qquad \tau_J > J|X_t = (n,\hat{n},k)] \nonumber\\
&= E[N_{t+JT} - N_{t}; \tau_J > J|X_t =(n,\hat{n},k)]\nonumber\\
&= \sum_{j = 0}^{J-1}E[d_T(N_{t+jT},\hat{N}_{t+jT}, K_{t+jT});
\tau_J
>
J|X_t =(n,\hat{n},k)]\nonumber\\
&\leq -\delta T J {\rm Pr}(\tau_J > J)\leq -\frac{\delta T J}{2}.
\end{align}
Combining \eqref{eq:drif_of_Lyapunov_unlikely} and
\eqref{eq:drift_of_Lyapunov_likely_case1}, we know that there exists
some $J$ such that inequation \eqref{eq:negative_drift_Lyapunov}
holds for some given $\Delta > 0$.

Now we are able to use the results about the hitting time bounds
implied by drift analysis in \cite{Hajek1982DriftAnal}. Let
\begin{equation*} \label{eq:}
\Lambda = [M+(JT)^2]\max\{1, \frac{1+3\gamma}{3\gamma},
\frac{\beta-3\gamma}{1-\beta + 3\gamma}\}.
\end{equation*}
Note that for any $K_t \in \mathbb{S}_K$, whenever $N_t \geq
M+(JT)^2$ or $\hat{N}_t \geq M+(JT)^2$,
\eqref{eq:negative_drift_Lyapunov} holds. According to Theorem 2.3
in \cite{Hajek1982DriftAnal}, for any initiate state, the returning
time $\tau^*_{\Lambda} = \min\{t > 0: V(N_t, \hat{N}_t, K_t) <
\Lambda\}$ is exponential type, which implies that $X_t$ is
geometrically ergodic and concludes the proof of Theorem
\ref{thm:stability_of_fasa}.
\end{proof} 

Similarly to the discussion in \cite{Tsitsiklis1987StabAnal}, from
the proof of Theorem \ref{thm:stability_of_fasa}, we know that the
proposed FASA scheme is stable under more general traffics, as long
as the average arrival rate $\bar{\lambda} < e^{-1}$ and the traffic
model satisfies the conditions in \eqref{eq:arr_process_exp_decay_1}
and \eqref{eq:arr_process_exp_decay_2}.

\section{Simulation results}\label{sec:sim_res}
In this section we evaluate the performance of the proposed scheme
through simulation. We first examine the tracking performance and
the effect of control parameters $\nu$ and $\eta$ on the access
performance. We then study the access delay of the proposed scheme,
including both the cases of single event and multiple events
reporting.

We compare the performance of our FASA scheme,
the ideal policy with perfect knowledge of backlog, PB-ALOHA
\cite{Rivest1987PBALOHA}, and Q$^+$-Algorithm \cite{Lee2007QPlus}. With perfect knowledge of $N_t$, the ideal policy sets transmission probability at $p_t = 1/N_t$ for $N_t > 0$. Thus, the ideal policy achieves the minimum access delay of S-ALOHA and serves as a benchmark in the comparison. For PB-ALOHA, we use the estimated arrival rate
$\hat{\lambda}_t = e^{-1}$, as suggested in
\cite{Tsitsiklis1987StabAnal}. Q$^+$-Algorithm belongs to the class of
multiplicative schemes which is first proposed by Hajek and van Loon
\cite{Hajek1982ITAC}. In Q$^+$-Algorithm, $\hat{N}_t$ is updated as
follows:
\begin{equation*} \label{eq:est_scheme_qplus}
\hat{N}_{t+1} = \max\{1, [I(Z_t = 0)/\zeta_0 + I(Z_t = 1) +\zeta_c
I(Z_t = c)] \hat{N}_{t}\}, \nonumber
\end{equation*}
where $\zeta_0 = 2^{0.25} \approx 1.1892$ and $\zeta_c = 2^{0.35}
\approx 1.2746$ are suggested in \cite{Lee2007QPlus} for optimal
performance.

\subsection{Performance of tracking}
In order to gain more insights into the operation of the estimate
schemes, we treat adaptive S-ALOHA schemes as dynamic systems
and study their step responses, where the number of backlogged
devices $N_t = 0$ when $t < 0$ and $N_t = n$ for all $t \geq 0$. We
examine the tracking performance of schemes for $n = 500$, $1000$,
and $2000$. For the sake of simplicity, we fix the value of $k_m$ in
FASA at 20, since the effect of $k_m$ vanishes when it is large
enough due to the the exponential decay of distribution of
$K_{t,c}$.


Before quantitive analysis, we first show the evolution of
estimations under different conditions, which gives us some
perceptual understanding about the behavior of these schemes.

%

\begin{figure}[thbp]
\begin{center}
\subfigure[$K_t$]{
\includegraphics[angle = 0,width = 0.8\linewidth]{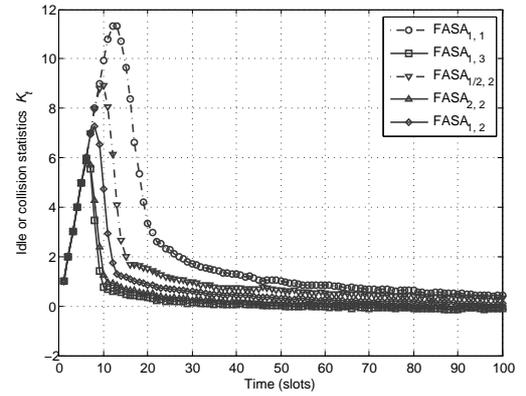}
\label{fig:aver_evol_kt_eta_nu} } \subfigure[$\hat{N}_t$]{
\includegraphics[angle = 0,width = 0.8\linewidth]{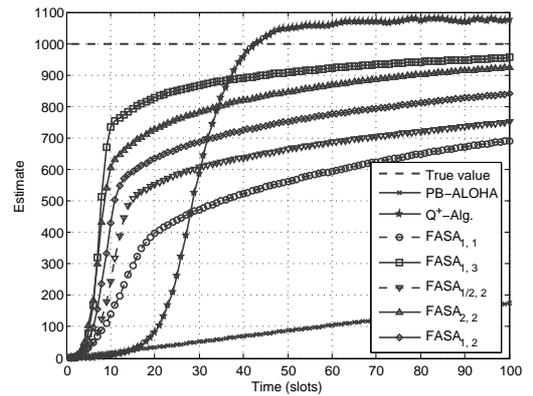}
\label{fig:aver_evol_estimate_eta_nu}}
 \caption{Evolution of
estimation with different parameters ($n = 1000$).}
\label{fig:aver_evol_eta_nu}
\end{center}
\end{figure}

Fig.~\ref{fig:aver_evol_eta_nu} shows the evolution of $K_t$ and
$\hat{N}_t$ for different adaptive schemes, where the subscripts of
FASA represents values of $\eta$ and $\nu$. Average values of $K_t$
and $\hat{N}_t$ in each slot are obtained from 4000 independent
experiments. It can be seen from this figure that, unlike the almost
linearly increasing in PB-ALOHA, the estimate $\hat{N}_t$ given by
FASA increases slowly at the beginning, but speeds up due to the
consecutive collisions, i.e., the increment of $K_t$. When the
estimate gets close to the true value, success and idle slots occur
and hence the increment of the estimate slows down. The estimate of
Q$^+$-Algorithm follows the same trend as FASA and speeds up even
faster on average than FASA because of the exponentially increment.
Though the drift of the estimate has the right direction,
Q$^+$-Algorithm turns out to be a bias estimate scheme, which will
result in the suffering of the throughput at steady state. When
comparing the curves of estimate for FASA with different parameters,
we can see that with larger $\eta$ or $\nu$, the estimate adjusts
faster.

\begin{figure}[thbp]
\begin{center}
\subfigure[$K_t$]{
\includegraphics[angle = 0,width = 0.8\linewidth]{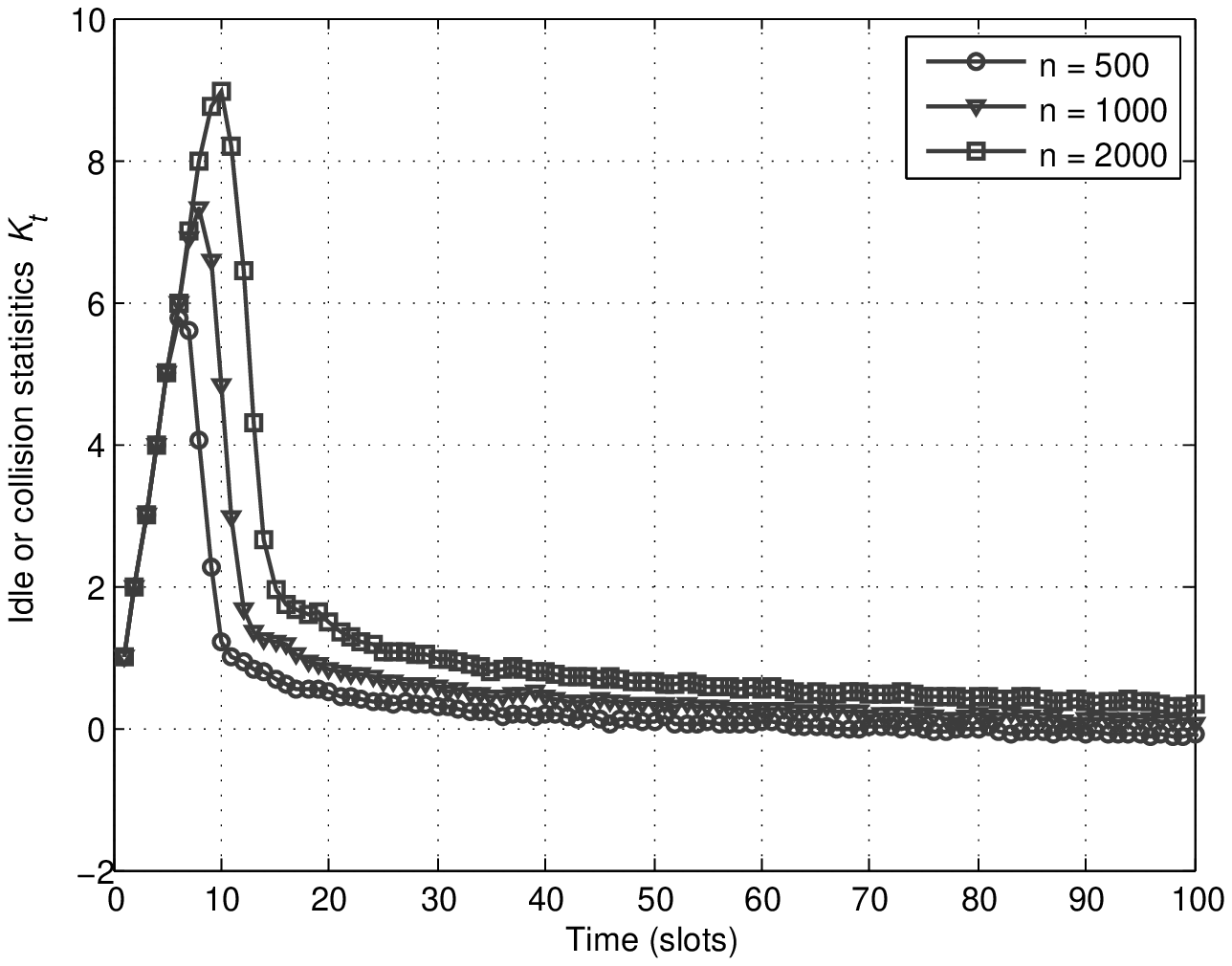}
\label{fig:aver_evol_kt_n} } \subfigure[$\hat{N}_t$]{
\includegraphics[angle = 0,width = 0.8\linewidth]{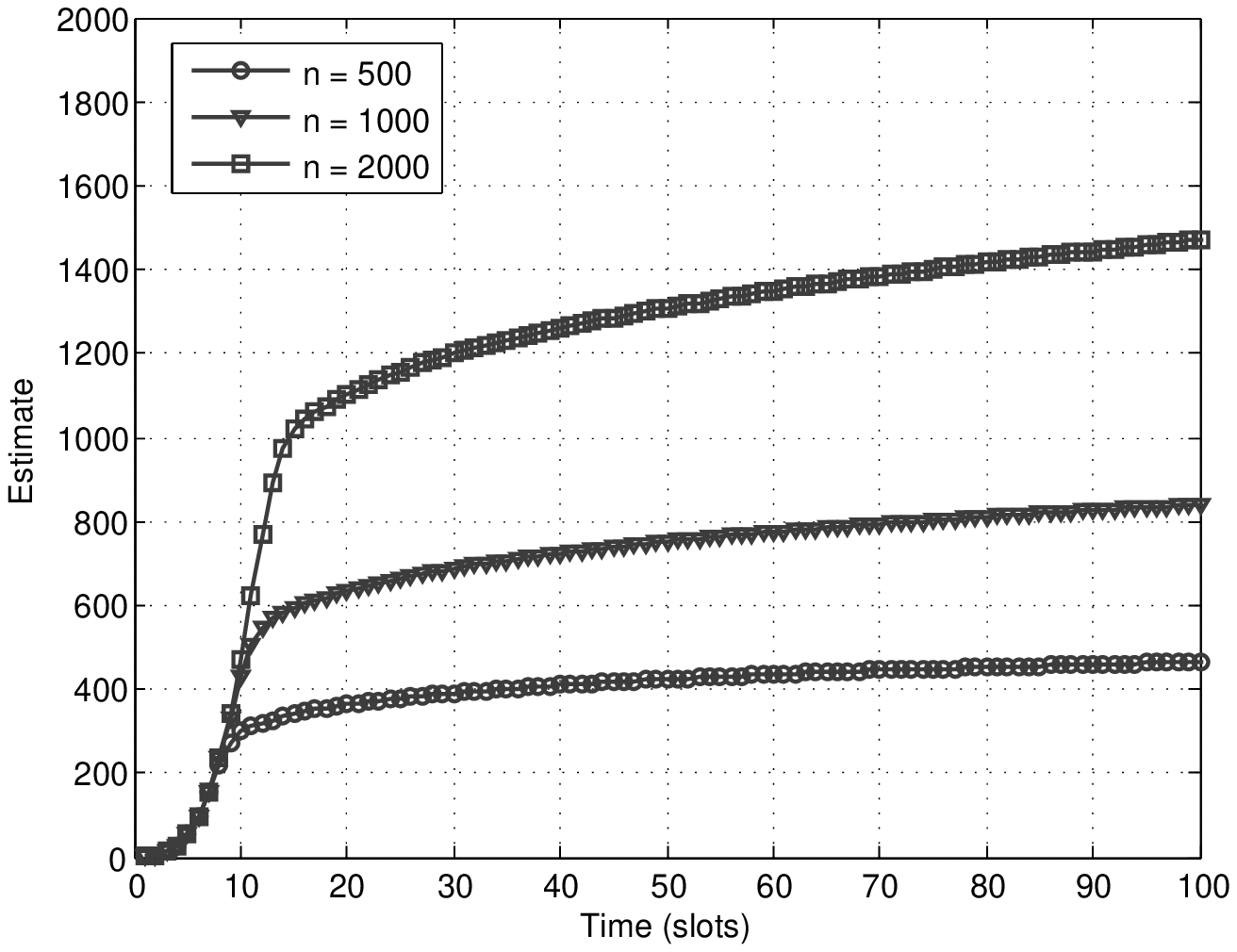}
\label{fig:aver_evol_estimate_n}} \caption{Evolution of FASA for
different $n$ ($\eta = 1, \nu = 2$).} \label{fig:aver_evol_n}
\end{center}
\end{figure}

Fig.~\ref{fig:aver_evol_n} shows the evolution of $K_t$ and
$\hat{N}_t$ for different numbers of devices $n$. As shown in the
figure, for larger $n$, there are more consecutive collisions at the
beginning, which results in larger increment of estimate. After the
peak point, the average value of $K_t$ mainly depends on the offered
load $\rho_t = n/\hat{N}_t$ or the ratio of $\hat{N_t}/n$, so does
the step size. For example, since $\hat{N}_{10} \approx 300$ for $n
= 500$, and $\hat{N}_{30} \approx 1200$ for $n = 2000$, i.e., the
ratios of $\hat{N_t}/n$ in these slots for $n = 500$ and 1000 are
both about 0.6, they have almost the same average value of $K_t$,
which is about 1. Though $K_t$ decreases and FASA behaves like a
fixed step size scheme as the estimate gets close to the true value,
the value of $K_t$ with larger $n$ is larger than that with small
$n$. Hence, we can expect that the time taken by FASA to catch up
the true value or certain proportion of the true value will increase
more slowly than linearly in $n$.

%


Because the access delay depends on the throughput, two
throughput-oriented metrics are introduced to measure the response
speed and stationary performance: 0\%-$x$\% throughput rising time
and stationary throughput. The 0\%-$x$\% throughput rising time is
defined as the time required for the expected throughput to rise
from 0\% to $x$\% of the optimal value $e^{-1}$. For $x = 10, 50$,
and $90$, they are equal to the time required for the estimated
number of backlogged devices $\hat{N}_t$ to rise from 0\% to
20.45\%, 37.34\%, and 65.25\% of the true value $n$, respectively.
Stationary throughput is the average throughput after the time that
the expected throughput reaches 90\% of the optimal value.


As shown in Table \ref{tab:rise_time}, for the same $x$, the
0\%-$x$\% rising time (unit: slot) of PB-ALOHA almost linearly increases in $n$ and is much
larger than that of Q$^+$-Algorithm and FASA. For instance, when $n
= 1000$, the 0\%-50\% rising time of FASA with $\eta = 1$ and $\nu =
2$ is about 1/20 of that of PB-ALOHA. Moreover, due to the
aggressive update in FASA, the 0\%-$x$\% rising time increases more
slowly rather than linearly as the number of devices $n$ increases,
especially when $x$ is small. Comparing the rising time of FASA with
different $\eta$ and $\nu$, we observe that the increment of $\eta$
or $\nu$ results in reduction of rising time. With multiplicative
adjustment, it takes longer time than FASA for Q$^+$-Algorithm to
reach 10\% of the optimal value but shorter time to increase the
expected throughput from 10\% to 90\% of the optimal value. In
addition, the increment of the rising time in Q$^+$-Algorithm is
tiny as $n$ grows. However, aggressive adjustment of estimate
usually results in large fluctuation at the steady state, and thus
lower stationary throughput, which is shown in Table
\ref{tab:stationary_throughput}. Thus, trade-off between the rising
time and stationary throughput is necessary for choosing the values
of parameters.

\begin{table}[!h]
\tabcolsep 0pt \caption{0\% - $x$\% throughput rising time}
\vspace*{-25pt}
\begin{center}
\def\temptablewidth{0.5\textwidth}
{\rule{\temptablewidth}{1pt}}
\begin{tabular*}{\temptablewidth}{@{\extracolsep{\fill}}ccccccccc}
$n$ &$x$ & {\fontsize{0.09in}{.08in}\selectfont{PB-ALOHA}} &
{\fontsize{0.09in}{.08in}\selectfont{Q$^+$-Alg.}}
 &{\fontsize{0.09in}{.08in}\selectfont{FASA$_{1,1}$}} &{\fontsize{0.09in}{.08in}\selectfont{FASA$_{1,3}$}}
 &{\fontsize{0.09in}{.08in}\selectfont{FASA$_{1,2}$}} &{\fontsize{0.09in}{.08in}\selectfont{FASA$_{1/2,2}$}}
 &{\fontsize{0.09in}{.08in}\selectfont{FASA$_{2,2}$}}\\
\hline
   &10& 59.8  & 21.0 & 9.1 & 6.0 & 7.1  &8.1 & 5.0\\
500   &50& 117.9  & 23.5 & 14.1 & 8.0 & 9.1  &11.8 & 7.5\\
   &90& 313.1  & 27.5 & 43.9 & 14.5& 21.3  &31.9 & 15.6\\\hline
   &10   &118.5 & 23.0  & 13.2 & 7.1 & 8.3  &10.1 & 7.1 \\
1000 &50   &237.0 & 26.6  & 21.6 & 9.4 & 12.0  &15.0 & 9.0 \\
   &90   &627.4 & 30.7  & 83.2 & 24.8 & 38.1  &60.0 & 20.6 \\\hline
   &10 & 236.0 & 26.0 & 18.3 & 8.1 & 10.1 &12.1 & 8.1\\
2000  &50 & 473.8 & 29.5 & 33.7 & 10.3 & 14.9 &19.7 & 11.9\\
   &90 & 1254.5 & 33.9 & 165.1 & 33.2 & 62.5 &115.1 & 38.1
\end{tabular*}
{\rule{\temptablewidth}{1pt}}
\end{center}
\label{tab:rise_time}
\end{table}

\begin{table}[!h]
\tabcolsep 0pt \caption{Stationary throughput} \vspace*{-25pt}
\begin{center}
\def\temptablewidth{0.5\textwidth}
{\rule{\temptablewidth}{1pt}}
\begin{tabular*}{\temptablewidth}{@{\extracolsep{\fill}}cccccccc}
$n$  & {\fontsize{0.09in}{.08in}\selectfont{PB-ALOHA}} &
{\fontsize{0.09in}{.08in}\selectfont{Q$^+$-Alg.}}
 &{\fontsize{0.09in}{.08in}\selectfont{FASA$_{1,1}$}} &{\fontsize{0.09in}{.08in}\selectfont{FASA$_{1,3}$}}
 &{\fontsize{0.09in}{.08in}\selectfont{FASA$_{1,2}$}} &{\fontsize{0.09in}{.08in}\selectfont{FASA$_{1/2,2}$}}
 &{\fontsize{0.09in}{.08in}\selectfont{FASA$_{2,2}$}}\\
\hline
500   &0.3686  & 0.3529 & 0.3679 & 0.3656 & 0.3670  &0.3676 & 0.3653\\
1000    &0.3684 & 0.3521 &0.3678 & 0.3663 & 0.3675   &0.3681 & 0.3668\\
2000   &0.3681 & 0.3523 &0.3678 & 0.3674 & 0.3677 &0.3678 & 0.3675
\end{tabular*}
{\rule{\temptablewidth}{1pt}}
\end{center}
\label{tab:stationary_throughput}
\end{table}

\subsection{Access delay}
In this section, simulation results about the access delay of
adaptive S-ALOHA schemes are presented. In some event-driven M2M
applications, response can be taken with partial messages from the
 detecting devices and not all devices need to report an event. Thus, both the
distribution of access delay for single event reporting and the long
term average delay for repetitive event reporting are evaluated to
study the performance of the proposed scheme.

\subsubsection{Single event reporting}

We focus on the scenarios where a large amount of devices are
activated to report a single event and study the distribution of
access delay of different adaptive schemes. Assume that $N_0 = n$
devices are triggered at the same time when an event is detected,
and attempt to access the BS on the RACH. The scenarios with the
number of active devices $n = 500$, $1000$, and $2000$ are studied.

Table \ref{tab:access_delay} provides the $y$\% access delay (unit: slot), which
is the access delay achieved by $y$\% of the active devices, and $y$
is set to $10$, $50$, and $90$. From Table \ref{tab:access_delay},
we can see that the performance of the proposed FASA scheme is close
to the benchmark with perfect information. For PB-ALOHA, it takes a
long time to track the number of backlogged devices and few devices
can access successfully during this period. For instance, the 10\%
delay is about two times of that for other schemes. For example,
when $n = 1000$, the 10\% delay of FASA$_{1,2}$ is 290.8 slots while
it is 542.4 slots for PB-ALOHA. With multiplicative increment, the
Q$^+$-Algorithm can track the number of backlogs in a short time
because of the exponential increment due to the consecutive
collision slots. However, it takes longer for all the devices to
access the channel under Q$^+$-Algorithm than under FASA due to the
large estimation fluctuations in
Q$^+$-Algorithm. Comparing the access delay achieved by FASA with
different $\eta$ and $\nu$, we observe that the 10\% access delay is
slightly smaller for larger $\eta$ or $\nu$, since they provide a
quicker response ability. However, the larger fluctuation makes the
50\% and 90\% access delay for larger $\eta$ and $\nu$ close to, or
even larger than that for smaller $\eta$ and $\nu$.

\begin{table}[!h]
\tabcolsep 0pt \caption{$y$\% Access Delay} \vspace*{-25pt}
\begin{center}
\def\temptablewidth{0.5\textwidth}
{\rule{\temptablewidth}{1pt}}
\begin{tabular*}{\temptablewidth}{@{\extracolsep{\fill}}cccccccccc}
$n$ &$y$ & {\fontsize{0.09in}{.08in}\selectfont{Perf.Info.}} &
{\fontsize{0.09in}{.08in}\selectfont{PB-ALOHA}} &
{\fontsize{0.09in}{.08in}\selectfont{Q$^+$-Alg.}}
 &{\fontsize{0.09in}{.08in}\selectfont{FASA$_{1,1}$}} &{\fontsize{0.09in}{.08in}\selectfont{FASA$_{1,3}$}}
 &{\fontsize{0.09in}{.08in}\selectfont{FASA$_{1,2}$}} &{\fontsize{0.09in}{.08in}\selectfont{FASA$_{1/2,2}$}}
 &{\fontsize{0.09in}{.08in}\selectfont{FASA$_{2,2}$}}\\
\hline
   &10 & 136.1 & 271.9  & 151.2 & 155.2 & 146.2 & 146.4  &152.9 & 143.1\\
500   &50& 680.6 & 822.3  & 712.5 & 702.7 & 694.4 & 691.3  &700.6 & 691.0\\
   &90 & 1223.1 & 1433.0  & 1282.9 & 1250.7 & 1252.1 & 1244.0  &1254.2 & 1247.8\\\hline
   &10 & 267.3  &542.4 & 298.6  & 306.4 & 285.9 & 290.8  &303.3 & 282.9 \\
1000 &50 & 1351.9  &1648.2 & 1426.0  & 1394.3 & 1384.7 & 1385.7  &1385.8 & 1378.5\\
   &90 & 2440.8  &2871.4 & 2558.7  & 2496.0 & 2488.2 & 2484.1  &2489.9 & 2483.5 \\\hline
   &10 & 545.3 & 1083.8 & 585.1 & 614.5 & 563.7 & 568.6 &592.4 & 564.3\\
2000  &50 & 2712.1 & 3294.9 & 2855.8 & 2793.4 & 2749.2 & 2752.1 &2779.1 & 2744.6\\
   &90 & 4886.3 & 5745.8 & 5124.3 & 4983.1 & 4944.9 & 4944.0 &4968.6 &
   4934.5
\end{tabular*}
{\rule{\temptablewidth}{1pt}}
\end{center}
\label{tab:access_delay}
\end{table}


\subsubsection{Repetitive events reporting with interrupted Poisson traffic}

The events happen sequentially in the real system and we now study
the long term average delay of adaptive schemes under interrupted
Poisson traffic with different arrival rates and bursty level. In
addition to average delay, we also define the normalized divergence
as follows to quantify the divergence from the theoretical optimum
performance:
\begin{equation} \label{eq:}
e(D) = \frac{D-D^*}{D^*},\nonumber
\end{equation}
where $D$ is the average delay of a particular scheme and $D^*$ is
the theoretical optimal delay with perfect information. As pointed
in the single event reporting case, the performance of FASA with
different $\eta$ and $\nu$ are rather close. Thus, only the
performance of FASA with $\eta = 1$ and $\nu = 2$ is presented here.

\begin{figure}[thbp]
\begin{center}
\subfigure[Average delay ($\theta = 0.0001$)]{
\includegraphics[angle = 0,width = 0.8\linewidth]{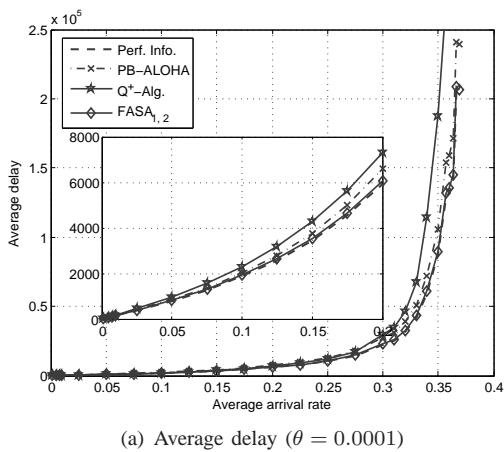}
\label{fig:aver_delay_diff_lambda_theta00001} }
\subfigure[Divergence ($\theta = 0.0001$)]{
\includegraphics[angle = 0,width = 0.8\linewidth]{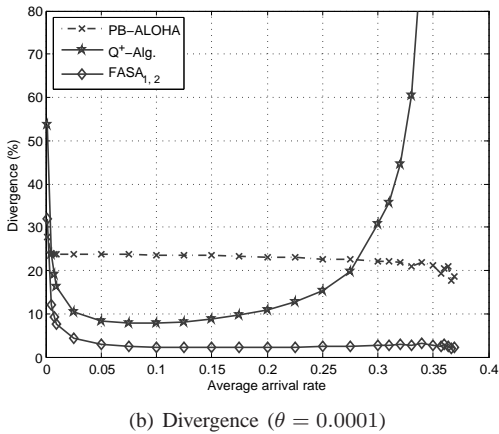}
\label{fig:aver_delay_divergence_diff_lambda_theta00001}}
\caption{Average access delay with different arrival rates.}
\label{fig:aver_delay_divergence_diff_lambda}
\end{center}
\end{figure}

Fig.~\ref{fig:aver_delay_divergence_diff_lambda} compares the
average delay and the normalized divergence of adaptive schemes
under different arrival rates and fixed ON-probability $\theta =
0.0001$. From Fig.~\ref{fig:aver_delay_diff_lambda_theta00001}, we
observe that both PB-ALOHA and FASA are stable when the average
arrival rate $\bar{\lambda} < e^{-1}$ and experience finite access
delays. For Q$^+$-Algorithm, however, when the arrival rate is
larger than about 0.352, the access delay grows unbounded,
indicating that the algorithm with the given parameters is unstable
for some $\bar{\lambda} < e^{-1}$. As pointed out in
\cite{Hajek1982DriftAnal}, the parameters in Q$^+$-Algorithm should
be carefully chosen to stabilize the scheme according to the value
of $\bar{\lambda}$, which is not required in either PB-ALOHA or FASA. As shown in
Fig.~\ref{fig:aver_delay_divergence_diff_lambda_theta00001}, when
the arrival rate is close to zero, the divergence of Q$^+$-Algorithm
and FASA are larger than that of PB-ALOHA due to the fluctuation of
estimation, while all the delays are very small. As the average
arrival rate increases, the divergence of FASA decreases and gets
close to the optimal value, since the estimate error becomes
relatively smaller compared to the increasing number of backlogged
devices in the system. For $\bar{\lambda}$ larger than 0.1, The
divergence of FASA is about 2.5\%, while it is about 22\% for
PB-ALOHA.

We point out that since the average access delay could be dominated
by the time waiting in the system after the estimate catches up the
true value, the improvement of performance by FASA does not seem to
be significant from the average delay point of view. However, as
discussed in the single event reporting cases, the 10\% access delay
can be improved significantly by FASA, which is very important to
the event-driven M2M communications.

\begin{figure}[thbp]
\begin{center}
\subfigure[$\bar{\lambda} = 0.05$]{
\includegraphics[angle = 0,width = 0.8\linewidth]{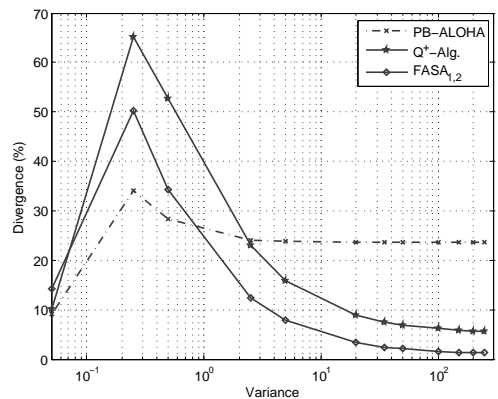}
\label{fig:aver_delay_divergence_diff_variance_lambda005}}
\subfigure[$\bar{\lambda} = 0.35$]{
\includegraphics[angle = 0,width = 0.8\linewidth]{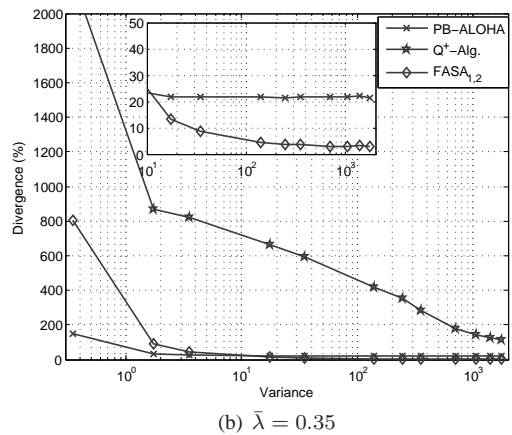}
\label{fig:aver_delay_divergence_diff_variance_lambda035} }
\caption{Divergence of average access delay under different
burstiness.} \label{fig:aver_delay_divergence_diff_variance}
\end{center}
\end{figure}

In order to examine the impact of burstiness, we present in Fig.
\ref{fig:aver_delay_divergence_diff_variance} the divergence of
access delay versus variance of the arrival process $\sigma_A^2$ for
$\bar{\lambda} = 0.05$ and $0.35$.  For the light traffic scenarios
with $\bar{\lambda} = 0.05$, when the bursty level is low, i.e., the
variance is small, the access delay obtained from all these scheme
are close to the optimal value. The reason is that there is usually
only one device is triggered in one slot when the estimate is
usually set to 1 after several idle slots. As the traffic becomes
more bursty, the divergence first increases owed to the rising time
and fluctuation of estimate; and then the divergence of FASA and
Q$^+$-Algorithm decreases for high bursty traffic because with
aggressive update, they are able to track the status of the network
quickly while the fluctuation become relatively smaller compared to
the total number of backlogs. For the high traffic scenarios with
$\bar{\lambda} = 0.35$, the divergences keep decreasing as the
bursty level increase, while the proposed FASA scheme performs better
than both PB-ALOHA and Q$^+$-Algorithm under high bursty traffic.
Since the traffic in the event-driven M2M applications is bursty, we
believe that our proposed scheme will perform well for these
applications.

\section{Conclusions}\label{sec:conclusions}
In this paper, we proposed a FASA scheme for
event-driven M2M communications. By adjusting the estimate of the
backlogs with statistics of consecutive idle and collision slots,
a BS can track the number of backlogged devices more quickly. That is a main advantage
compared to fixed step size additive schemes, e.g., PB-ALOHA.
Moreover, we studied the stability of the proposed FASA under bursty
traffic, which is modeled as an interrupted Poisson process. By
analyzing the $T$-slot drifts of the FASA, we showed that without
modifying the values of parameters, the proposed FASA scheme is
stable for any average arrival rate less than $e^{-1}$, in the sense
that the system is geometrically ergodic. This property results in a
much better long term average performance under heavy traffic loads,
as compared with that of multiplicative schemes. In summary, the
proposed scheme is an effective and stable S-ALOHA scheme and is
suitable for the random access control of event-driven M2M
communications as well as other systems characterized by bursty
traffic.

\appendices

\section{Proof of Lemma \ref{thm:prob_approx}}\label{app:proof_of_prob_approx}
Recall that for $\Delta n = \Delta \hat{n} = 0$, it has been proved
in \cite{Hajek1982ITAC} that when either the number of backlogged
devices $n$ or its estimate $\hat{n}$ is sufficiently large, the
idle and success probabilities (and hence as well as the collision
probability) can be approximated as functions of the offered load
$\rho = n/\hat{n}$. We generalize the results to any given $\Delta
n$ and $\Delta \hat{n}$ by showing that as $n$ or $\hat{n}$ tends to
infinity, the difference between the offered loads $(n + \Delta
n)/(\hat{n} + \Delta \hat{n})$ and $\rho$ can be ignored and the
distribution of access results can still be approximated by the same
functions of $\rho$.

First, consider inequation (\ref{eq:idle_prob_gapbound}), which is
used for approximating the probability of $Z_t = 0$. Notice that
\begin{align*} \label{eq:approx_idle}
G^{(0)}&=\left|e^{-\frac{n}{\hat{n}}} - (1 - \frac{1}{\hat{n} +
\Delta \hat{n}})^{n
+ \Delta n}\right|\nonumber\\
& \leq \underset{G_1^{(0)}}{\underbrace{\left|e^{-\frac{n}{\hat{n}}}
- e^{-\frac{n + \Delta n}{\hat{n} + \Delta \hat{n}}}\right|}}  +
\underset{G_2^{(0)}}{\underbrace{\left|e^{-\frac{n + \Delta
n}{\hat{n} + \Delta \hat{n}}} - (1 - \frac{1}{\hat{n} + \Delta
\hat{n}})^{n + \Delta n}\right|}}.\nonumber
\end{align*}

For $G_2^{(0)}$, according to Proposition 2.1 in
\cite{Hajek1982ITAC}, we know that there exists some $M_2^{(0)} >
0$, such that $G_2^{(0)}\leq \epsilon/2$ for any $(n, \hat{n}) \in
H_{M_2^{(0)}}$, where $H_{M_2^{(0)}} = \{(n, \hat{n}): n \geq
M_2^{(0)} \mbox{ or } \hat{n} \geq M_2^{(0)}, n + \Delta n \geq 0,
\hat{n} + \Delta \hat{n} \geq 1\}$. Hence, we just need to show that
$G_1^{(0)} \leq \epsilon/2$ when either $n$ or $\hat{n}$ is
sufficiently large.

Since $\lim_{\rho \to \infty} e^{-\rho} = 0$, there exists some
$\hat{\rho}^{(0)} > 1$ such that $e^{-\rho} \leq \epsilon/4$ for any
$\rho > \hat{\rho}^{(0)}$. Given a number $\rho^{(0)}
> \hat{\rho}^{(0)}$, we analyze the value of $G_1^{(0)}$ when $n/\hat{n} > \rho^{(0)}$ and $0 < n/\hat{n} \leq
\rho^{(0)}$, respectively.

\subsubsection*{0-a) $n/\hat{n} > \rho^{(0)}$}

When $n/\hat{n} > \rho^{(0)} > \hat{\rho}^{(0)}$, it is easy to show
that there exists some $M_{1,1}^{(0)}
> 0$, such that if $n \geq
M_{1,1}^{(0)}$, then $(n + \Delta n)/(\hat{n} + \Delta \hat{n}) \geq
(n - |\Delta n|)/(\hat{n} + |\Delta \hat{n}|) > \hat{\rho}^{(0)}$.
Hence, when $n/\hat{n} > \rho^{(0)}$ and $n \geq M_{1,1}^{(0)}$, we
have
\begin{eqnarray*} \label{eq:}
G_1^{(0)} \leq e^{-\frac{n}{\hat{n}}} + e^{-\frac{n + \Delta
n}{\hat{n} + \Delta \hat{n}}} \leq  \epsilon/4 + \epsilon/4 =
\epsilon/2. \nonumber
\end{eqnarray*}

\subsubsection*{0-b) $0 < n/\hat{n} \leq \rho^{(0)}$}

Since $n/\hat{n} \geq 0$, we have
\begin{align*} \label{eq:}
G_1^{(0)} & =  e^{-\frac{n}{\hat{n}}} \left|1 -
\exp(\frac{n}{\hat{n}}-\frac{n +
\Delta n}{\hat{n} + \Delta \hat{n}})\right|\nonumber\\
& \leq  \left|1 - \exp \left[\frac{n \Delta \hat{n} - \hat{n} \Delta
n}{\hat{n}(\hat{n} + \Delta \hat{n})}\right] \right|.\nonumber
\end{align*}

When $0 \leq n/\hat{n} \leq \rho^{(0)}$, we have
\begin{eqnarray*} \label{eq:}
\left|\frac{n \Delta \hat{n} - \hat{n} \Delta n}{\hat{n}(\hat{n} +
\Delta \hat{n})}\right| \leq \frac{\rho^{(0)} |\Delta \hat{n}| + |
\Delta n|}{(\hat{n} + \Delta \hat{n})} \to 0,\nonumber
\end{eqnarray*}
as $\hat{n}\to \infty$. Therefore, $\lim_{\hat{n} \to \infty}
\exp[\frac{n \Delta \hat{n} - \hat{n} \Delta n}{\hat{n}(\hat{n} +
\Delta \hat{n})}] = 1$, and hence, there exists some $M_{1,
2}^{(0)}$, such that $G_1^{(0)} \leq \epsilon/2$ for any
$(n,\hat{n})$ satisfying $0 \leq n/\hat{n}\leq \rho^{(0)}$ and
$\hat{n}\geq M_{1,2}^{(0)}$.


Consequently, combining all the cases analyzed above and letting
$M^{(0)} = \max\{M_2^{(0)}, M_{1,1}^{(0)}, \rho^{(0)}
M_{1,2}^{(0)}\}$ follows that $G^{(0)} \leq G_1^{(0)} +
G_2^{(0)}\leq \epsilon$ for any $(n, \hat{n}) \in H_{M^{(0)}}$.

Next, we turn to the proof of inequation
(\ref{eq:succ_prob_gapbound}), which is used for approximating
probability of $Z_t = 1$. Similarly to inequation
(\ref{eq:idle_prob_gapbound}),
\begin{align*} \label{eq:approx_succ}
G^{(1)} &= \left|\frac{n}{\hat{n}}e^{-\frac{n}{\hat{n}}} -
\frac{n+\Delta n}{\hat{n} + \Delta \hat{n}}(1 - \frac{1}{\hat{n} +
\Delta \hat{n}})^{n +
\Delta n - 1}\right|\nonumber \\
& \leq
\underset{G_1^{(1)}}{\underbrace{\left|\frac{n}{\hat{n}}e^{-\frac{n}{\hat{n}}}
- \frac{n + \Delta n}{\hat{n} + \Delta \hat{n}}e^{-\frac{n + \Delta
n}{\hat{n} + \Delta
\hat{n}}}\right|}} \nonumber \\
&+  \underset{G_2^{(1)}}{\underbrace{\left|\frac{n + \Delta
n}{\hat{n} + \Delta \hat{n}}e^{-\frac{n + \Delta n}{\hat{n} + \Delta
\hat{n}}} -\frac{n+\Delta n}{\hat{n} + \Delta \hat{n}}(1 -
\frac{1}{\hat{n} + \Delta \hat{n}})^{n + \Delta n -
1}\right|}}.\nonumber
\end{align*}

Using the result in \cite{Hajek1982ITAC}, we know that there exists
some $M_2^{(1)} > 0$, such that $G_2^{(1)}\leq \epsilon/2$ for any
$(n, \hat{n}) \in H_{M_2^{(1)}}$ and we only need to show that
$G_2^{(1)}\leq \epsilon/2$ under certain conditions.

Since $\lim_{\rho \to 0}\rho e^{-\rho}  = \lim_{\rho \to \infty}\rho
e^{-\rho} = 0$, there exist some $\hat{\rho}_1^{(1)}$ and
$\hat{\rho}_2^{(1)}$, such that $0< \hat{\rho}_1^{(1)} < 1 <
\hat{\rho}_2^{(1)}$, and $\rho e^{-\rho} \leq \epsilon/4$ for any
$\rho$ satisfying $0 < \rho < \hat{\rho}_1^{(1)}$ or $\rho
> \hat{\rho}_2^{(1)}$. Given $\rho_1^{(1)} \in (0,
\hat{\rho}_1^{(1)})$ and $\rho_2^{(1)} \in (\hat{\rho}_2^{(1)},
\infty)$, we study the following three cases based on the range of
$n/\hat{n}$:

\subsubsection*{1-a) $0 < n/\hat{n} < \rho_1^{(1)}$}

Similarly to the analysis of 0-{\it a}), since $\rho_1^{(1)} <
\hat{\rho}_1^{(1)}$, there exists some $M_{1,1}^{(1)}
> 0$, such that if $\hat{n} > M_{1,1}^{(1)}$, then $(n + \Delta n)/(\hat{n} + \Delta \hat{n}) <
\hat{\rho}_1^{(1)}$ and hence
\begin{align*} \label{eq:}
G_1^{(1)} \leq \frac{n}{\hat{n}} e^{-\frac{n}{\hat{n}}} + \frac{n +
\Delta n}{\hat{n} + \Delta \hat{n}} e^{-\frac{n + \Delta n}{\hat{n}
+ \Delta \hat{n}}} \leq \epsilon / 2.\nonumber
\end{align*}

\subsubsection*{1-b) $n/\hat{n}> \rho_2^{(1)}$}

Similarly to cases 0-{\it a} and 1-{\it a}), we can show that there
exists some $M_{1,2}^{(1)} > 0$, such that if $n > M_{1,2}^{(1)}$,
then $G_1^{(1)} \leq \epsilon /2$.

\subsubsection*{1-c) $\rho_1^{(1)} \leq n/\hat{n} \leq \rho_2^{(1)}$}

When $\rho_1^{(1)}\leq n/\hat{n} \leq \rho_2^{(1)}$, we have
\begin{align*} \label{eq:}
G_1^{(1)}& = \frac{n}{\hat{n}}e^{-n/\hat{n}} \left| 1-
\frac{\hat{n}}{n} \cdot \frac{n+\Delta n}{\hat{n} + \Delta
\hat{n}}\exp\left[\frac{n \Delta \hat{n} - \hat{n} \Delta
n}{\hat{n}(\hat{n} + \Delta \hat{n})}\right]\right|\nonumber\\
&\leq  \rho_2^{(1)}\left| 1- \frac{1+\Delta n / n}{1 + \Delta
\hat{n} / \hat{n}}\exp\left[\frac{n \Delta \hat{n} - \hat{n} \Delta
n}{\hat{n}(\hat{n} + \Delta \hat{n})}\right]\right|.\nonumber
\end{align*}

Since
\begin{eqnarray} \label{eq:}
\left|\frac{1+\Delta n / n}{1 + \Delta \hat{n} / \hat{n}}-1\right|
\leq \frac{|\Delta n| / \rho_1^{(1)} + |\Delta \hat{n}|}{|\hat{n} +
\Delta \hat{n}|} \to 0, \nonumber
\end{eqnarray}
as $\hat{n} \to \infty$, we have $\lim_{\hat{n} \to \infty}
\frac{1+\Delta n / n}{1 + \Delta \hat{n} / \hat{n}} = 1$ and
$\lim_{\hat{n} \to \infty} \frac{1+\Delta n / n}{1 + \Delta \hat{n}
/ \hat{n}}\exp[\frac{n \Delta \hat{n} - \hat{n} \Delta
n}{\hat{n}(\hat{n} + \Delta \hat{n})}] = 1$. Hence, there exists a
$M_{1, 3}^{(1)}$ such that $G_1^{(1)} \leq \epsilon/2$ for any
$(n,\hat{n})$ satisfying $\rho_1^{(1)}  \leq n/\hat{n}\leq
\rho_2^{(1)}$ and $\hat{n} \geq M_{1,3}^{(1)}$.

Therefore, $G^{(1)} \leq G_1^{(1)} + G_2^{(1)}\leq \epsilon$ for any
$(n, \hat{n}) \in H_{M^{(1)}}$, where
\begin{eqnarray} \label{eq:}
M^{(1)} = \max\{M_2^{(1)}, M_{1,1}^{(1)}, M_{1,2}^{(1)},
\rho_2^{(1)} M_{1,3}^{(1)}\}.\nonumber
\end{eqnarray}

Finally, from the above analysis, choosing $M = \max\{M^{(0)},
M^{(1)}\}$, we know that (\ref{eq:idle_prob_gapbound}) and
(\ref{eq:succ_prob_gapbound}) hold for any $(n, \hat{n}) \in H_M$
and this concludes the proof of Lemma \ref{thm:prob_approx}.

\section{Proof of Theorem \ref{thm:equilibrium_of_fasa}}\label{app:proof_of_equilibrium_of_fasa}
The proposition can be proved by calculating the derivative of $\varphi(\rho)$.

For given values of $\nu$ and $k_m$, let
\begin{align}
\varphi^{(0)}(\rho) &= |q_0(\rho) E[\Delta N_t |(0, n, \hat{n})]|
\nonumber\\
&= q_0(\rho)[1 + h_0(\nu)\mu(\nu, q_0(\rho),k_m)],\nonumber
\end{align}
and
\begin{align}
\varphi^{(c)}(\rho) &= |q_c(\rho)E[\Delta N_t | (c, n, \hat{n})]|\nonumber\\
&= q_c(\rho)[(e-2)^{-1} + h_c(\nu)\mu(\nu, q_c(\rho),k_m)].\nonumber
\end{align}

Then $\varphi^{(0)}(1) = \varphi^{(c)}(1) = e^{-1} + \eta$ and
$\varphi(1) = -\varphi^{(0)}(1) + \varphi^{(c)}(1) = 0$. Next, we
claim that, for given $\nu >0$, $\mu(\nu, q, k_m)$ defined in
(\ref{eq:nu_order_moment}) is an increasing function of $q$ ($0 < q
< 1$). This is because
\begin{align} \label{eq:}
\frac{\partial \mu}{\partial q}&=\sum_{k = 1}^{k_{m}-1} k^\nu q^{k-2}[k(1-q)-1] + (k_{m}-1)k_{m}^\nu q^{k_{m} -2} \nonumber\\
&=  \sum_{k = 1}^{k^*} k^\nu q^{k-2}[k(1-q)-1]\nonumber\\
&\quad + \sum_{k = k^*+1}^{k_{m}-1} k^\nu q^{k-2}[k(1-q)-1] + (k_{m}-1)k_{m}^\nu q^{k_{m} -2}\nonumber \\
&>(k^*)^\nu \left\{\sum_{k = 1}^{k_{m}-1} q^{k-2}[k(1-q)-1] +
(k_{m}-1) q^{k_{m} -2}\right\} \nonumber\\
&= (k^*)^\nu \left[\sum_{k = 1}^{k_{m}} (k-1)q^{k-2} - \sum_{k =
1}^{k_{m}-1} k q^{k-1} \right]= 0, \nonumber
\end{align}
where $k^* = \lfloor (1-q)^{-1} \rfloor$ is the largest integer not
greater than $(1-q)^{-1}$, and thus $k^\nu \leq (k^*)^\nu$ if $1\leq
k \leq k^*$ and $k^\nu > (k^*)^\nu$ if $k > k^*$. In addition, the
idle probability $q_0(\rho) = e^{-\rho}$ is nonnegative and strictly
decreasing in $\rho$. Hence, $\mu(\nu, q_0(\rho),k_m)$ is strictly
decreasing in $\rho$ and $\varphi^{(0)}(\rho)$ is a strictly
decreasing function of $\rho$. On the other hand, since $q_c(\rho) =
1-e^{-\rho} -\rho e^{-\rho}$ is nonnegative and strictly increasing
in $\rho$, we can similarly show that $\varphi^{(c)}(\rho)$ is a
strictly increasing function of $\rho$. Thus, $\varphi(\rho) =
-\varphi^{(0)}(\rho) + \varphi^{(c)}(\rho)$ is a strictly increasing
function of $\rho$. Consequently, $\varphi(\rho) < \varphi(1) =0$
when $0 < \rho < 1$ and $\varphi(\rho)
>\varphi(1) =0$ when $\rho
> 1$.

\section{Proof of Lemma
\ref{thm:approx_drift}}\label{app:proof_of_approx_drift}

Because of the similarity, we present a complete analysis of
inequation \eqref{eq:approx_drift_gap_n}, while discuss briefly
about inequation \eqref{eq:approx_drift_gap_nhat} at the end. Recall
that we assume the same realization of arrival process $A_{t+s}$ for
$X_{t+s}$ and $X'_{t+s}$. In addition, given $T$ and $(n,\hat{n},
k)$, the number of departures between slot $t$ and $t+T$ is bounded
and the number of new arrivals can also be bounded with high
probability. In order to use Lemma \ref{thm:prob_approx}, we
consider separately the events $\sum_{s = 0}^{T-1} A_{t + s}
> B_A$ and $\sum_{s = 0}^{T-1} A_{t+s}\leq B_A$. \\

{\it 1) $\sum_{s = 0}^{T-1} A_{t + s}> B_A$}

Using Chernoff bound, we can show that the probability ${\rm
Pr}(\sum_{s = 0}^{T-1} A_{t + s} > B_A)$ decays exponentially as
$B_A$ grows to infinity. With the assumption of same realization of
$A_{t+s}$, the differences between the $T$-slot drift of $N_{t+s}$
and $N'_{t+s}$ is bounded by $T$, i.e., $|d_T(n, \hat{n}, k) -
d'_T(n, \hat{n}, k)|\leq T$.
Therefore, for any given $\epsilon > 0$, there exists some $B_A > 0$
such that
\begin{align} \label{eq:approx_drift_gap_n_unlikely}
|d_T(n, \hat{n}, k) - d'_T(n, \hat{n}, k)|{\rm Pr}(\sum_{s =
0}^{T-1} A_{t + s} > B_A) \leq \epsilon/2.
\end{align}

{\it 2) $\sum_{s = 0}^{T-1} A_{t + s} \leq B_A$}

Let $\mathbf{A}_{t,T} = (A_t, A_{t+1}, \ldots, A_{t+T-1})$,
$\mathbf{Z}_{t,T} = (Z_t, Z_{t+1}, \ldots, Z_{t+T-1})$, and
$\mathbf{Z}'_{t,T} = (Z'_t, Z'_{t+1}, \ldots, Z'_{t+T-1})$. The set
of possible values of $(\mathbf{A}_{t,T}, \mathbf{Z}_{t,T})$ and
$(\mathbf{A}_{t,T}, \mathbf{Z}'_{t,T})$ are the same, denoted by
$\Omega$. Since $Z_{t+s}$ or $Z'_{t+s}$ has three possible values
and $\sum_{s = 0}^{T-1} A_{t + s} \leq B_A$, $\Omega$ is a finite
set. Each pair $(\mathbf{a}, \mathbf{z}) = (a_0 \ldots a_{T-1}, z_0
\ldots z_{T-1})\in \Omega$ results in corresponding drifts in both
$N_{t+s}$ and $N'_{t+s}$, denoted by $\Delta N_{(\mathbf{a},
\mathbf{z}, n,\hat{n},k)}$ and $\Delta N'_{(\mathbf{a}, \mathbf{z},
n,\hat{n},k)}$, respectively.

One of the differences between $X_{t+s}$ and $X'_{t+s}$ is that we
do not limit the values of $N'_{t+s}$ and $\hat{N}'_{t+s}$, i.e.,
$N'_{t+s} < 0$ and $\hat{N}'_{t+s} < 1$ are allowed in the virtual
sequence, which is not the case in $X_{t+s}$. However, noticing the
fact that when $n \geq T$, the drifts of $N_{t+s}$ and $N'_{t+s}$
are only decided by $(\mathbf{a}, \mathbf{z})$, we have $\Delta
N_{(\mathbf{a}, \mathbf{z},n,\hat{n},k)} = \Delta N'_{(\mathbf{a},
\mathbf{z},n,\hat{n},k)}$ for any $(\mathbf{a}, \mathbf{z})\in
\Omega$ in this case. We first study the case where $n \geq T$, and
analyze later the other cases where $n$ is not large enough.

Conditioned on $\sum_{s = 0}^{T-1} A_{t + s} \leq B_A$, we define
the following probabilities:
\begin{align*}
&\quad f_{\mathbf{A}}(\mathbf{a}) = {\rm Pr}(\mathbf{A}_{t,T} =
\mathbf{a}),\nonumber\\
&\quad f_{(\mathbf{A}, \mathbf{Z})|X}(\mathbf{a},
\mathbf{z}|n,\hat{n},k)\\
&= {\rm Pr}[(\mathbf{A}_{t,T} ,\mathbf{Z}_{t,T}) = (\mathbf{a},
\mathbf{z})|X_t = (n,\hat{n},k)], \nonumber\\
&\quad f_{(\mathbf{A}, \mathbf{Z}')|X'}(\mathbf{a},
\mathbf{z}|n,\hat{n},k) \\
&= {\rm Pr}[(\mathbf{A}_{t,T} ,\mathbf{Z}'_{t,T}) = (\mathbf{a},
\mathbf{z})|X'_t = (n,\hat{n},k)], \nonumber\\
&\quad f_{\mathbf{Z}|\mathbf{A},X}(\mathbf{z}|\mathbf{a},
n,\hat{n},k) \\
&= {\rm Pr}[\mathbf{Z}_{t,T} =
 \mathbf{z}|\mathbf{A}_{t,T} = \mathbf{a},X_t = (n,\hat{n},k)],
 \nonumber\\
&\quad f_{\mathbf{Z}'|\mathbf{A},X'}(\mathbf{z}|\mathbf{a},
n,\hat{n},k) \\
&= {\rm Pr}[\mathbf{Z}'_{t,T} =
 \mathbf{z}|\mathbf{A}_{t,T} = \mathbf{a}, X'_t = (n,\hat{n},k)]. \nonumber
\end{align*}

Thus, when $n \geq T$, we have
\begin{align*}\label{eq:}
&\quad |d_T(n, \hat{n}, k) - d'_T(n, \hat{n}, k)|{\rm Pr}(\sum_{s =
0}^{T-1}
A_{t + s} \leq B_A)\nonumber\\
&\leq \big|\sum_{(\mathbf{a}, \mathbf{z})\in \Omega}[f_{(\mathbf{A},
\mathbf{Z})|X}(\mathbf{a}, \mathbf{z}|n,\hat{n},k)\\
&\quad -f_{(\mathbf{A}, \mathbf{Z}'|X'}(\mathbf{a},
\mathbf{z}|n,\hat{n},k)] \Delta
N_{(\mathbf{a}, \mathbf{z},n,\hat{n},k)}\big|\nonumber\\
&= \big|\sum_{(\mathbf{a}, \mathbf{z})\in
\Omega}[f_{\mathbf{Z}|\mathbf{A},X}(\mathbf{z}|\mathbf{a},n,\hat{n},k)\\
&\quad
-f_{\mathbf{Z}'|\mathbf{A},X}(\mathbf{z}|\mathbf{a},n,\hat{n},k)]
f_{\mathbf{A}}(\mathbf{a})\Delta
N_{(\mathbf{a}, \mathbf{z},n,\hat{n},k)}\big|\nonumber\\
&\leq C \sum_{(\mathbf{a}, \mathbf{z})\in
\Omega}\left|f_{\mathbf{Z}|\mathbf{A},X}(\mathbf{z}|\mathbf{a},
n,\hat{n},k)- f_{\mathbf{Z}'|\mathbf{A},X}(\mathbf{z}|\mathbf{a},
n,\hat{n},k)\right|,
\end{align*}
where $C$ is the maximum value of $|f_{\mathbf{A}}(\mathbf{a})\Delta
N_{(\mathbf{a}, \mathbf{z},n,\hat{n},k)}|$ for all $(\mathbf{a},
\mathbf{z})\in \Omega$ and $k \in \{-k_m, -k_m + 1, \ldots, 0,
\ldots, k_m-1, k_m\}$.

Letting
\begin{align*}
&g_{Z_s}(z_s|\mathbf{a},z_0 \ldots z_{s-1})={\rm Pr}[Z_{t+s} =
 z_s|X_t = (n,\hat{n},k),\nonumber\\
&\qquad \qquad \mathbf{A}_{t,T} = \mathbf{a}, Z_t \ldots Z_{t+s-1} =
z_0 \ldots z_{s-1}],
 \nonumber\\
&g_{Z'_s}(z_s|\mathbf{a},z_t \ldots z_{t+s-1})= {\rm Pr}[Z'_{t+s} =
z_s|X'_t = (n,\hat{n},k),\nonumber\\
&\qquad \qquad \mathbf{A}_{t,T} = \mathbf{a}, Z'_t \ldots Z'_{t+s-1}
= z_0 \ldots z_{s-1}], \nonumber
\end{align*}
we have
\begin{align*}
&\quad f_{\mathbf{Z}|\mathbf{A},X}(\mathbf{z}|\mathbf{a}, n,
\hat{n},k) \\
&= {\rm Pr}(Z_t = z_0)\prod_{s = 1}^{T-1} g_{Z_s}(z_s|\mathbf{a},z_0
\ldots z_{s-1})
\nonumber\\
&=  \prod_{s=0}^{T-1} {\rm Pr}[Z_{t+s} = z_s|(N_{t+s},
\hat{N}_{t+s})=(n_s,\hat{n}_s)_{\mathbf{a}, \mathbf{z},
n,\hat{n},k}],\nonumber
\end{align*}
\begin{align*}
&\quad f_{\mathbf{Z}'|\mathbf{A},X}(\mathbf{z}|\mathbf{a}, n,
\hat{n},k) \\
&= {\rm Pr}(Z'_t = z_0)\prod_{s = 1}^{T-1}
g_{Z'_s}(z_s|\mathbf{a},z_0 \ldots z_{s-1})
\nonumber\\
&= \prod_{s=0}^{T-1} {\rm Pr}[Z'_{t+s} = z_s|(N_{t+s},
\hat{N}'_{t+s})=(n_s,\hat{n}_s)_{\mathbf{a}, \mathbf{z},
n,\hat{n},k}],\nonumber
\end{align*}
where $(n_s,\hat{n}_s)_{\mathbf{a}, \mathbf{z}, n,\hat{n},k}$ is the
number of backlogged devices and its estimate in slot $t+s$, given
the initial state $X_t = X'_t = (n,\hat{n}, k)$, $\mathbf{A}_{t,T}
=\mathbf{a}$, and $\mathbf{Z}_{t,T} = \mathbf{Z}'_{t,T} =
\mathbf{z}$.

According to the construction of $X'_{t+s}$, the distribution of
$Z'_{t+s}$ is fixed, i.e., ${\rm Pr}(Z'_{t+s} = i) = q_i(\rho)$;
while for $X_{t+s}$, there are finite possible values of $N_{t+s}
-n$ and $\hat{N}_{t+s} - \hat{n}$, for all $s = 1, 2, \ldots, T-1$,
since the initial value of $K_t$ is from a finite set. Therefore,
using Lemma \ref{thm:prob_approx}, we know that there exists some
sufficiently large $M'_1$, such that if $n \geq M'_1$ or $\hat{n}
\geq M'_1$, then for all $s = 1,2, \ldots, T-1$, and $(\mathbf{a},
\mathbf{z})\in \Omega$,
\begin{align} \label{eq:}
&\left|{\rm Pr}[Z_{t+s} = z_s|(N_{t+s},
\hat{N}_{t+s})=(n_s,\hat{n}_s)_{\mathbf{a},
\mathbf{z}, n,\hat{n},k}]\right.\nonumber\\
&\left.- {\rm Pr}[Z'_{t+s} = z_s|(N_{t+s},
\hat{N}'_{t+s})=(n_s,\hat{n}_s)_{\mathbf{a}, \mathbf{z},
n,\hat{n},k}]\right| \leq \frac{\epsilon}{2 T C|\Omega|},\nonumber
\end{align}
where $|\Omega|$ is the number of elements in $\Omega$. Hence
\begin{equation*} \label{eq:}
\left|f_{\mathbf{Z}|\mathbf{A},X}(\mathbf{z}|\mathbf{a}, n,
\hat{n},k)-f_{\mathbf{Z}'|\mathbf{A},X}(\mathbf{z}|\mathbf{a}, n,
\hat{n},k)\right| \leq \frac{\epsilon}{2 C|\Omega|},\nonumber
\end{equation*}
and thus,
\begin{eqnarray} \label{eq:approx_drift_gap_n_likely}
|d_T(n, \hat{n}, k) - d'_T(n, \hat{n}, k)|{\rm Pr}(\sum_{s =
0}^{T-1} A_{t + s} \leq B_A) \leq \epsilon/2.
\end{eqnarray}

Therefore, combining \eqref{eq:approx_drift_gap_n_likely} with
\eqref{eq:approx_drift_gap_n_unlikely} implies that
\eqref{eq:approx_drift_gap_n} holds when $n \geq T$, and either $n
\geq M'_1$ or $\hat{n} \geq M'_1$.

Now consider the cases where $n < T$. In these cases,
$(\mathbf{A}_{t,T}, \mathbf{Z}_{t,T}) = (\mathbf{a},\mathbf{z})$ is
an impossible event for some $(\mathbf{a},\mathbf{z})\in \Omega$, if
it results in some $N_{t+s} < 0$. For these
$(\mathbf{a},\mathbf{z})$, we set $\Delta N_{(\mathbf{a},
\mathbf{z},n,\hat{n},k)} = \Delta N'_{(\mathbf{a},
\mathbf{z},n,\hat{n},k)}$, which does not affect the calculation of
drifts since the probabilities of these events in $X_{t+s}$ are
zero. Notice that for these $(\mathbf{a},\mathbf{z})$, there is at
least one component of $\mathbf{z}$ equal to 1. Because $\lim_{\rho
\to 0} \rho e^{-\rho} = 0$, we can find a number $\rho_1 \in (0,1)$,
such that $\rho e^{-\rho} \leq \epsilon /(2 T C|\Omega|)$ for all
$\rho = n/\hat{n} \leq \rho_1$. Then the difference between
$f_{\mathbf{Z}|\mathbf{A},X}(\mathbf{z}|\mathbf{a}, n, \hat{n},k)$
and $f_{\mathbf{Z}'|\mathbf{A},X}(\mathbf{z}|\mathbf{a}, n,
\hat{n},k)$ is bounded by $\epsilon /(2 T C|\Omega|)$ and the same
conclusion holds when $n < T$ and $n/\hat{n} \leq \rho_1$.

Consequently, from the analysis above, we know that inequation
\eqref{eq:approx_drift_gap_n} holds when $n \geq M_1$ or $\hat{n}
\geq M_1$, where $M_1 = \max\{T/\rho_1, M'_1/\rho_1\}$.

Finally, we discuss about inequation
\eqref{eq:approx_drift_gap_nhat}. For given $(\mathbf{a},
\mathbf{z}) \in \Omega$, denote the corresponding $T$-slot drifts of
$\hat{N}_{t+s}$ and $\hat{N}'_{t+s}$ by $\Delta
\hat{N}_{(\mathbf{a}, \mathbf{z}, n,\hat{n},k)}$ and $\Delta
\hat{N}'_{(\mathbf{a}, \mathbf{z}, n,\hat{n},k)}$, respectively.
Note that for inequation \eqref{eq:approx_drift_gap_n}, we do not
specially treat the case where $\hat{n}$ is not large enough (but
$n$ is large enough), since we still have $\Delta
\hat{N}_{(\mathbf{a}, \mathbf{z}, n,\hat{n},k)} = \Delta
\hat{N}'_{(\mathbf{a}, \mathbf{z}, n,\hat{n},k)}$ and the update of
FASA guarantees that $\hat{N}_{t+s} \geq 1$. However, $\Delta
\hat{N}_{(\mathbf{a}, \mathbf{z}, n,\hat{n},k)} \neq \Delta
\hat{N}'_{(\mathbf{a}, \mathbf{z}, n,\hat{n},k)}$ may occur for some
$(\mathbf{a}, \mathbf{z})$ when $\hat{n}$ is small. Since for all
$(\mathbf{a}, \mathbf{z})\in \Omega$, both $\Delta
\hat{N}_{(\mathbf{a}, \mathbf{z}, n,\hat{n},k)}$ and $\Delta
\hat{N}'_{(\mathbf{a}, \mathbf{z}, n,\hat{n},k)}$ are bounded
uniformly in $(n,\hat{n},k)$, the impact of $\Delta
\hat{N}_{(\mathbf{a}, \mathbf{z}, n,\hat{n},k)} \neq \Delta
\hat{N}'_{(\mathbf{a}, \mathbf{z}, n,\hat{n},k)}$ can be made
ignorable by making the probability of this event as close to zero
as possible with the fact that $\lim_{\rho \to \infty}e^{-\rho} =
0$. Hence, similarly to inequation \eqref{eq:approx_drift_gap_n}, we
can analyze the following three cases to obtain the threshold for
inequation \eqref{eq:approx_drift_gap_nhat}:

a) $n \geq T, \hat{n} \geq T[1 + k_m^\nu h_0(\nu)]+1$;

b) $n < T, \hat{n} \geq T[1 + k_m^\nu h_0(\nu)]+1$;

c) $n \geq T, \hat{n} < T[1 + k_m^\nu h_0(\nu)]+1$.

Therefore, the proof of this lemma can be concluded by choosing $M$
as the lager threshold for inequation \eqref{eq:approx_drift_gap_n}
and inequation \eqref{eq:approx_drift_gap_nhat}.

\section{Proof of Lemma
\ref{thm:prop_est_err_drift}}\label{app:proof_of_est_err_drift}

a) Let
\begin{align*} \label{eq:}
\psi^{(0)}(\rho, \bar{\lambda}) &= - e^{-\rho} +\frac{1}{e-2}(1-
e^{-\rho}-\rho e^{-\rho} ),\nonumber\\
\psi^{(1)}(\rho, \bar{\lambda}) &= \rho e^{-\rho} -
\bar{\lambda},\nonumber\\
\psi^{(2)}(\rho, \bar{\lambda})& = - e^{-\rho} h_0(\nu)\mu(\nu,
q_0(\rho),k_{m})\nonumber\\
&\quad +(1- e^{-\rho}-\rho e^{-\rho} )h_c(\nu)\mu(\nu,
q_c(\rho),k_{m}).\nonumber
\end{align*}
Then, $\psi = \psi^{(0)} + \psi^{(1)} + \psi^{(2)}$.

First, because
\begin{eqnarray} \label{eq:}
\frac{\partial(\psi^{(0)} + \psi^{(1)})}{\partial \rho} & =
&e^{-\rho} + \frac{\rho e^{-\rho}}{e-2} +e^{-\rho} -\rho e^{-\rho}>0
,\nonumber
\end{eqnarray}
$\psi^{(0)} + \psi^{(1)}$ is strictly increasing in $\rho$.

In addition, we have shown in Appendix \ref{app:proof_of_equilibrium_of_fasa} that
$\mu(\nu, q_0(\rho), k_{m})$ is strictly decreasing in $\rho$ while
$\mu(\nu, q_c(\rho), k_{m})$ is strictly increasing in $\rho$. Thus
it is easy to verify that $\psi^{(2)}(\rho, \bar{\lambda})$ is
strictly increasing in $\rho$. Consequently, $\psi = \psi^{(0)} +
\psi^{(1)} + \psi^{(2)}$ is strictly increasing in $\rho$.

b) For any $\bar{\lambda} \in (0, e^{-1}]$, we have $\psi(1,
\bar{\lambda}) = e^{-1} - \bar{\lambda} \geq 0$, and $\psi(\rho,
\bar{\lambda}) \to -h_0(\nu)(k_m)^{\nu +1} < 0$ as $\rho \to 0$. In
addition, the function $\psi$ is continuous and strictly monotonic
in $\rho$. Thus, there is a unique solution $\rho =
\omega(\bar{\lambda}) \in (0,1]$ for equation $\psi(\rho,
\bar{\lambda}) = 0$.

c) For given $\bar{\lambda} \in (0, e^{-1})$, $\psi^{(0)}$ and
$\psi^{(2)}$ are both strictly increasing in $\rho$. In addition,
$\psi^{(0)} = \psi^{(2)} = 0$ when $\rho = 1$. Because the solution
$\rho = \omega(\bar{\lambda}) < 1$, we have
$\psi^{(0)}(\omega(\bar{\lambda}), \bar{\lambda}) +
\psi^{(2)}(\omega(\bar{\lambda}), \bar{\lambda})< 0$ and thus
$\psi^{(1)}(\omega(\bar{\lambda}), \bar{\lambda}) =
\omega(\bar{\lambda}) e^{-\varphi(\bar{\lambda})} - \bar{\lambda} >
0$, i.e., $\omega(\bar{\lambda}) e^{-\omega(\bar{\lambda})} >
\bar{\lambda}$.

\section{Proof of Lemma
\ref{thm:strictly_non_zero_drift}}\label{app:proof_of_strictly_non_zero_drift}
Since the $T$-slot drifts of $X_t$ can be approximated by the drifts
of $X'_{t+s}$, which can be further approximated by closed form
expressions, we start our proof by examining the properties of these
expressions and choose the values of $\gamma$ and $\delta$. Then by
choosing $T$ and $M$ properly, we make the approximate errors small
enough such that the actual drifts have the same properties as their
approximations.

Notice that $\bar{\lambda} - \rho e^{-\rho} < 0$ when $\rho = 1$ or
$\rho = \beta$. In addition, it is a continuous function and is
monotonically increasing in $\rho \in [\beta, 1]$. Therefore, there
exist some $\delta_1$ and $\gamma$ such that $\bar{\lambda} - \rho
e^{-\rho} < \delta_1$ for all $\rho \in [\beta - 5\gamma, 1+
5\gamma]$. For $\Psi(\rho, \bar{\lambda})$, with its strict
monotonicity in $\rho$, we conclude that $\psi(\rho, \bar{\lambda})
< \psi(\beta -\gamma, \bar{\lambda}) < 0$ for all $\rho \in (0,\beta
-\gamma)$ and $\psi(\rho, \bar{\lambda})
> \psi(1 + \gamma, \bar{\lambda}) > 0$ for all $\rho \in (1+\gamma,
\infty)$. Thus, there exists some $\delta$ such that
\begin{align}
&\bar{\lambda} - \rho e^{-\rho} \leq -3\delta/2,&\vee \rho &\in
[\beta - 5\gamma, 1+ 5\gamma],\label{eq:negative_drift_approx_n}\\
&\psi(\rho, \bar{\lambda})  \leq -3\delta, & \vee  \rho &\in (0,\beta -\gamma),\label{eq:negative_drift_approx_n_tilde}\\
&\psi(\rho, \bar{\lambda})  \geq 3\delta, & \vee  \rho &\in
(1+\gamma, \infty).\label{eq:positive_drift_approx_n_tilde}
\end{align}

Now, fix $\gamma$ and $\delta$. Using the uniform convergence of
$\frac{1}{T}\tilde{d}'_T(n, \hat{n}, k)$, we can choose a
sufficiently large $T$, such that for any $(n,\hat{n},k)\in
\mathbb{S}_X$,
\begin{equation}\label{eq:bound_approx_drift_n_tilde}
\left|\frac{1}{T}\tilde{d}'_T(n, \hat{n}, k) - \psi(\rho,
\bar{\lambda})\right| \leq \delta.
\end{equation}
where $\rho = n/\hat{n}$.

According to Lemma \ref{thm:approx_drift}, with the chosen $T$,
there exists some $M > 0$, such that if $n \geq M$ or $\hat{n} \geq
M$, then
\begin{align}
&\left|\frac{1}{T}d_T(n, \hat{n}, k) - \frac{1}{T}d'_T(n, \hat{n},
k)\right|\leq \delta/2,\label{eq:bound_drift_n}\\
&\left|\frac{1}{T}\hat{d}_T(n, \hat{n}, k) -
\frac{1}{T}\hat{d}'_T(n, \hat{n}, k)\right|\leq \delta/2,
\end{align}
and thus,
\begin{align}\label{eq:bound_drift_n_tilde}
\left|\frac{1}{T}\tilde{d}_T(n, \hat{n}, k) -
\frac{1}{T}\tilde{d}'_T(n, \hat{n}, k)\right| \leq \delta.
\end{align}

With $\frac{1}{T}d'_T(n, \hat{n}, k) = \bar{\lambda} - \rho
e^{-\rho}$, \eqref{eq:negative_drift_approx_n} and
\eqref{eq:bound_drift_n} together imply that for any $(n, \hat{n},
k) \in S_{5\gamma, M}$,
\begin{equation*}\label{eq:}
\frac{1}{T}d_T(n, \hat{n}, k) \leq -\delta,
\end{equation*}
and thus \eqref{eq:negative_drift_n} holds.

Similarly, \eqref{eq:negative_drift_n_tilde} follows by combining
\eqref{eq:negative_drift_approx_n_tilde},
\eqref{eq:bound_approx_drift_n_tilde}, and
\eqref{eq:bound_drift_n_tilde}; \eqref{eq:positive_drift_n_tilde}
follows by combining \eqref{eq:positive_drift_approx_n_tilde},
\eqref{eq:bound_approx_drift_n_tilde}, and
\eqref{eq:bound_drift_n_tilde}.

\bibliography{FA_ALOHA}

\end{document}